\documentclass[journal]{IEEEtran}
\usepackage{bbm}
\usepackage{cite}
\usepackage{amsfonts}
\usepackage{mathrsfs}
\usepackage{graphicx}
\usepackage{amssymb}
\usepackage{latexsym}
\usepackage{amsmath}
\usepackage{stfloats}
\usepackage{cases}
\usepackage{setspace}
\usepackage{bigstrut}
\usepackage{bm}
\usepackage{url}
\usepackage{color}
\usepackage{subfigure}
\usepackage[ruled,linesnumbered]{algorithm2e}
\usepackage{amsthm}
\usepackage{CJKutf8}

\newcommand{\bl}[1]{\color{blue}#1}

\begin{document}

\title{Analytical Framework of Airy Beams in Near-Field XL-MIMO: From Ideal Optics to Wireless Reality}
    
    \author{Zhizheng Lu,~\IEEEmembership{Student Member, IEEE}, Yu Han,~\IEEEmembership{Member, IEEE}, Ruirui Sun,~\IEEEmembership{Student Member, IEEE},
    
    Jiali Nie,~\IEEEmembership{Student Member, IEEE}, and Shi Jin,~\IEEEmembership{Fellow, IEEE}
    
    \thanks{Z. Lu, Y. Han, R. Sun, J. Nie, and S. Jin are with the School of Information Science and Engineering, Southeast University, Nanjing 210096, China (e-mail: luzz@seu.edu.cn; hanyu@seu.edu.cn; sunrr@seu.edu.cn; jialinie@seu.edu.cn; jinshi@seu.edu.cn).}
    }
\maketitle

\begin{abstract}

The synthesis of Airy-profiled wavefronts has emerged as a pivotal paradigm for advanced electromagnetic engineering, attributed to their intrinsic non-diffractive propagation, transverse self-acceleration, and structural self-healing properties. While the advent of extremely large-scale multiple-input multiple-output (XL-MIMO) and the elevation in frequency bands for sixth generation wireless systems provide the physical foundation for generating such structured beams, their wireless realization is fundamentally governed by hybrid precoding architectures, finite array apertures, and discrete antenna topologies. These constraints induce significant deviations from ideal optical Airy beam models, necessitating a rigorous re-characterization of Airy beams in practical wireless contexts. Consequently, this paper establishes an analytical theoretical framework to explicitly characterize Airy beam propagation in near-field XL-MIMO and derives the constraints on array aperture and antenna spacing to sustain distortion-free main lobe trajectories. Furthermore, quantitative metrics are developed to rigorously evaluate the performance trade-offs between Airy beams and Gaussian focusing beams, thereby providing systematic guidelines for their deployment in scenario-dependent wireless applications. Numerical results corroborate the proposed analytical theoretical framework of Airy beams in near-field XL-MIMO, and demonstrate the potential to achieve robust communication and spectral efficiency (SE) improvement in certain scenarios.

\textit{Index Terms}\textemdash Airy beams, analytical theoretical framework, near-field, robustness and SE enhancement, XL-MIMO.

\end{abstract}

\section{Introduction}\label{Sec: Introduction}

The evolution of extremely large-scale multiple-input multiple-output (XL-MIMO) systems is set to precipitate a technological revolution in the paradigm of sixth generation (6G) wireless systems \cite{XLMIMOSystems1}. The deployment of a substantial number of antennas at the base station (BS), together with the elevation of higher frequency bands, has been demonstrated to greatly enhance spectral efficiency (SE), area capacity, and coverage \cite{XLMIMOSystems2}. It is evident that the increase in array apertures and frequency bands leads to a near-field effect, thus enabling Gaussian beams to achieve significant power enhancement by near-field beam focusing \cite{XLMIMOSystems4}. However, the robustness may be compromised in highly dynamic scenarios and with positioning errors, while the SE undergoes a substantial reduction when the focusing position deviates from the actual position \cite{XLMIMOSystems8}. In addition, communication interruptions may occur under severe blockage, due to the straight-line propagation characteristic of Gaussian beams \cite{XLMIMOSystems9}.

Airy beams have attracted considerable attention within the domain of optics \cite{AiryBeam1, AiryBeam2, AiryBeam3}. Compared to Gaussian beams, Airy beams exhibit unique properties\cite{AiryBeam4, AiryBeam5, AiryBeam6}. \textit{1)} Non-diffraction: The transverse intensity profile remains approximately invariant during propagation, exhibiting strong resistance to diffraction-induced spreading; \textit{2)} Self-acceleration: The beam displays a parabolic trajectory during propagation, rather than a straight-line trajectory; \textit{3)} Self-healing: The original wavefront can be automatically reconstructed after being blocked by obstacles. These properties facilitate beam maintaining ability from near-field to far-field regions, programmable trajectories, and the ability to ‌circumvent obstacles, thus enhancing robustness and SE in wireless communication \cite{AiryBeam7}.

Many studies have been reported on the construction of Airy beams in sub-terahertz (THz) near-field regimes, which explored the potential of curved trajectories to circumvent obstacles \cite{AiryBeam8, AiryBeam9, AiryBeam10}. However, the construction of Airy beams in practical XL-MIMO systems is challenging due to the limited number of antennas, finite transmit power, and deployment of hybrid precoding architectures. The authors in \cite{Works1} conducted theoretical and experimental research on the construction of an Airy beam in the microwave frequency bands using an array equipped with microstrip patch antenna elements. In the work \cite{Works2}, Airy beams were constructed using THz multiple-input multiple-output (MIMO) systems, which added cubic phases to the antennas by analog phase shifters. Subsequently, a physics-informed neural network was developed for the purpose of ‌circumventing obstacles. In the work \cite{Works3}, a truncated Airy beam was constructed in microwave frequency bands by antenna arrays, thus improving wireless power propagation efficiency. In addition, the authors in \cite{Works4} applied Airy beams in quasi-line-of-sight scenarios for wireless data centers, and subsequently derived the main lobe trajectories of Airy beams in THz near-field communication \cite{Works5}.

The considerably larger array apertures and higher frequency bands facilitate the potential to construct Airy beams in XL-MIMO systems for near-field communication (i.e., the cubic phase method and the cubic phase plus focusing phase method \cite{Works2}). Nevertheless, this will also lead to fundamental changes in beam properties, such as the aberrant electric field (EF) and main lobe trajectories, as well as the non-negligible ``truncation'' and ``sampling'' errors (i.e., from an infinite and continuous plane to a finite and discrete array). Existing approaches for Airy beam construction primarily rely on exhaustive scanning \cite{Scanning}. Although a main lobe trajectory design scheme has been proposed in \cite{Works5}, constraints on array aperture and antenna spacing still require a rigorous quantitative metric. Consequently, a comprehensive exploration and analysis of the actual properties of Airy beams in near-field XL-MIMO is imperative. Moreover, theoretical verification is required to determine when Airy beams outperform Gaussian beams. The practical application of Airy beams in XL-MIMO systems for near-field communication remains constrained by a paucity of fundamental theoretical models.

In this paper, an analytical theoretical framework of Airy beams in XL-MIMO systems is established, while a quantitative comparison with Gaussian beams is developed. Moreover, scenario-dependent use cases of Airy beams in wireless communication are presented to enhance the robustness and SE. The main contributions are summarized as follows:

\begin{itemize}
\item \textbf{\textit{Analytical theoretic framework:}} An analytical theoretic framework of Airy beams in XL-MIMO systems is established, including the actual EFs and main lobe trajectories for cubic phase methods and cubic phase plus focusing phase methods. The numerical constraints on array aperture and antenna spacing, and the distortion-free main lobe trajectory range in wireless communication are also derived.
\end{itemize}

\begin{itemize}
\item \textbf{\textit{Comparison with Gaussian beams:}} A quantitative analysis for EF strength of Airy beams and Gaussian beams is developed in a scenario characterized by obstacles. Under minimal obstruction, the apparent obstacle avoidance of Airy beams stems only from main lobe deviation without power compensation effect. In instances of severe obstruction, Airy beams achieve superior performance that that of near-field focusing beams.
\end{itemize}

\begin{itemize}
\item \textbf{\textit{Scenario-dependent use cases:}} The scenario-dependent use cases of Airy beams in wireless communication are presented, which facilitates the potential of Airy beams for robust communication tools, and improves SE by serving multiple users by a single beam. Numerical results demonstrate that Airy beams can improve robustness in highly dynamic scenarios and with positioning errors, while achieving a trade-off between beamforming gain and served user number.
\end{itemize}

The rest of this paper is organized as follows: In Section $\rm \uppercase \expandafter{\romannumeral2}$, ideal Airy beams and the system model are introduced. In Section $\rm \uppercase \expandafter{\romannumeral3}$, an analytical theoretic framework of Airy beams in XL-MIMO systems is established. In Section $\rm \uppercase \expandafter{\romannumeral4}$, a comparison with Gaussian beams and scenario-dependent use cases are developed. Finally, conclusions are given in Section $\rm \uppercase \expandafter{\romannumeral5}$.

\textit{Notations:} Henceforth, lower-case and upper-case bold letters represent vectors and matrices, respectively. For a matrix $\bf A$, ${\bf A}^{\rm H}$ denotes the conjugate transpose operation; $\left| \cdot \right|$ denotes taking the absolute value operation, while $\min \left(\cdot\right)$ and $\max \left(\cdot \right)$ denote taking the minimum and maximum values.

\section{Airy Beams in XL-MIMO Systems}\label{Sec: System Model}

\subsection{Airy Beam Paradigm}

When transmitting electromagnetic waves from the $x-y$ plane (i.e., $z = 0$) and propagating in a free space in a directional manner along the $z$-axis (i.e., the paraxial propagation condition and for $z > 0$), the Helmholtz equation can be alternatively expressed as the paraxial wave equation \cite{References1}, such that
\begin{equation}\label{Eq: Section2_1_1}
j\frac{\partial \psi \left(x, y, z\right)}{\partial z} + \frac{1}{2k} \left( \frac{\partial^2 \psi \left(x, y, z\right)}{\partial x^2} + \frac{\partial^2 \psi \left(x, y, z\right)}{\partial y^2} \right) = 0,
\end{equation}
where $z \ge 0$, and $k = \frac{2\pi}{\lambda}$ denotes the wave number, with $\lambda$ being the wavelength; $\psi \left(x, y, z > 0\right)$ and $\psi \left(x, y, 0\right)$ represent the EF on the position $\left(x,y,z\right)$ and in the initial $x-y$ plane. Since the formula {\bl\eqref{Eq: Section2_1_1}} is symmetric to the parameters $x$ and $y$, it can then be assumed that \cite{References1}
\begin{equation}\label{Eq: Section2_1_2}
\psi \left(x, y, z\right) = \psi_x \left(x, z\right) \psi_y \left(y, z\right).
\end{equation}
Substituting {\bl\eqref{Eq: Section2_1_2}} into {\bl\eqref{Eq: Section2_1_1}}, such that
\begin{equation}\label{Eq: Section2_1_3}
\left\{ {\begin{array}{*{20}{c}}
{j\frac{\partial \psi_x \left(x, z\right)}{\partial z} + \frac{1}{2k} \frac{\partial^2 \psi_x \left(x, z\right)}{\partial x^2} = 0}\\
{j\frac{\partial \psi_y \left(y, z\right)}{\partial z} + \frac{1}{2k} \frac{\partial^2 \psi_y \left(y, z\right)}{\partial y^2} = 0}
\end{array}} \right..
\end{equation}
Due to the formula {\bl\eqref{Eq: Section2_1_1}} can be decoupled from the $x$ and the $y$ directions, only the EF in the $x$-$z$ plane is then analyzed.

In optics engineering, the Fourier transform is typically performed for the formula {\bl\eqref{Eq: Section2_1_3}} (which translates the spatial-domain EF, $\psi_{x} \left(x, z\right)$, into the frequency domain $k_{x}$, and indicates it as $\Psi \left(k_x, z\right)$) \cite{References2}, and thus we have
\begin{equation}\label{Eq: Section2_1_4}
\frac{\partial \Psi \left(k_x, z\right)}{\partial z} = -j \frac{k_x^2}{2k} \Psi \left(k_x, z\right),
\end{equation}
where $z \ge 0$, and
\begin{equation}\label{Eq: Section2_1_5}
\Psi \left(k_x, z\right) = \int_{-\infty}^{+\infty} {\psi_x \left(x,z\right) \exp\left[-jk_xx\right]} dx.
\end{equation}
Since the formula {\bl\eqref{Eq: Section2_1_4}} is an ordinary differential equation, the analytical solution of $\Psi \left(k_x, z\right)$ can be uniquely identified from the initial solution $\Psi \left(k_x, 0\right)$, for any $z > 0$, and denoted as
\begin{equation}\label{Eq: Section2_1_6}
\Psi \left(k_x, z\right) = \Psi \left(k_x, 0\right) \exp\left[-j \frac{k_x^2}{2k}z\right].
\end{equation}

\begin{figure}
  \centering
  \includegraphics[scale=0.34]{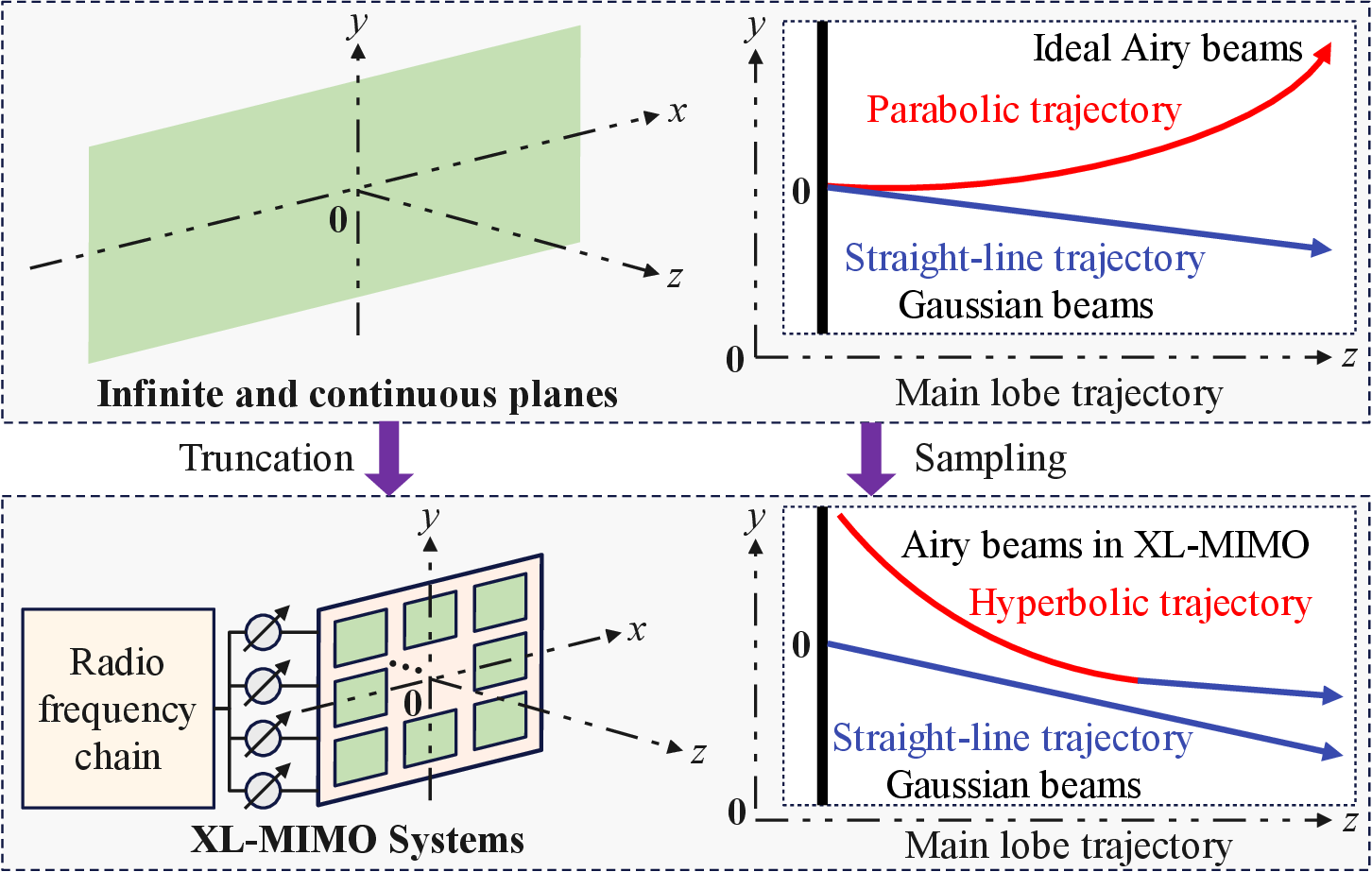}
  \caption{Airy and Gaussian beams in XL-MIMO systems.} \label{Fig: system model}
\end{figure}

In practical Airy beam engineering, the method of stationary phase (MSP) is adopted \cite{References3}, where $\Psi \left(k_x, 0\right)$ is set as
\begin{equation}\label{Eq: Section2_1_7}
\Psi \left(k_x, 0\right) = \exp \left[j \frac{x_0^3}{3} k_x^3\right],
\end{equation}
and $x_0 \ne 0$ is a transverse scale that controls the curvature of an Airy beam in the $x-z$ plane. Thus, the initial EF is
\begin{equation}\label{Eq: Section2_1_8}
\begin{aligned}
&\psi_x \left(x, 0\right) = \frac{1}{2\pi}\int_{-\infty}^{+\infty} {\Psi \left(k_x, 0\right) \exp\left[jk_xx\right]} d k_x \\ = &\frac{1}{2\pi}\int_{-\infty}^{+\infty} {\exp\left[j\frac{x_0^3}{3}k_x^3 + jk_x x\right]} d k_x = \frac{1}{x_0} {\rm Ai} \left(\frac{x}{x_0}\right),
\end{aligned}
\end{equation}
where ${\rm Ai} \left(x\right)$ denotes the Airy function, such that
\begin{equation}\label{Eq: Section2_1_9}
{\rm Ai} \left(x \right) = \frac{1}{2\pi}\int_{-\infty}^{+\infty} {\exp \left[j \left(\frac{t^3}{3} + xt\right)\right]} dt.
\end{equation}
Based on {\bl\eqref{Eq: Section2_1_6}} and {\bl\eqref{Eq: Section2_1_7}}, the EF on the position $\left(x,z\right)$ can be calculated according to \cite{Works3}, such that
\begin{equation}\label{Eq: Section2_1_10}
\begin{aligned}
&\psi_x \left(x, z\right) = \frac{1}{2\pi}\int_{-\infty}^{+\infty} {\Psi \left(k_x, z \right) \exp\left[j k_x x \right]} d k_x \\ = & \frac{1}{x_0} {\rm Ai} \left(\frac{x}{x_0} - \frac{z^2}{4k^2x_0^4} \right) \exp\left[j\left(\frac{xz}{2kx_0^3}- \frac{z^3}{12k^3x_0^6}\right)\right],
\end{aligned}
\end{equation}
for any $z > 0$. Subsequently, the main lobe trajectory of an ideal Airy beam can be calculated by
\begin{equation}\label{Eq: Section2_1_11}
\frac{x}{x_{\rm 0}} - \frac{z^2}{4k^2 x_0^4} = \mu_0,
\end{equation}
where $\mu_0 \approx -1.01879$ is the global maximum value of ${\rm Ai} \left(x\right)$, while ${\rm Ai} \left(\mu_0\right) \approx 0.5357$ \cite{AiryBeam5}. Consequently, the ideal main lobe trajectory can be expressed as \cite{Works3}
\begin{equation}\label{Eq: Section2_1_12}
x\left(z\right) = \frac{1}{4k x_0^3}z^2 + \mu_0 x_0.
\end{equation}
Note that it has a parabolic trajectory along $z$, while having a linear transverse trajectory velocity, $\frac{\partial x\left(z\right)}{\partial z} = \frac{1}{2k x_0^3}z$, and a constant transverse trajectory acceleration, $\frac{\partial^2 x\left(z\right)}{\partial z^2} = \frac{1}{2k x_0^3}$.

\subsection{XL-MIMO System Model}

The deployment of XL-MIMO significantly enlarges both the antenna count and array aperture, thereby enabling the construction of Airy beams in near-field communication. As shown in {\bl Fig.~\ref{Fig: system model}}, an XL-MIMO system with a low-cost hybrid precoding architecture is considered in this paper. The carrier frequency is indicated by $f_{\rm c}$, while the wavelength can be denoted as $\lambda = \frac{c}{f_{\rm c}}$, where $c$ is the speed of light. The BS is equipped with a uniform planar array (UPA) with $M$ rows and $N$ columns, and the spacing between adjacent antennas is indicated by $d$. The array apertures of each row and each column are indicated by $L_x = Nd$ and $L_y = Md$, respectively. The UPA is deployed along the $x$-$y$ plane, with the center of the UPA set as the origin, while the antennas in each row and each column are employed along the $x$-axis and the $y$-axis, respectively. A radio frequency chain is connected to all the antennas in the UPA, with an individual analog phase shifter followed by each antenna.

Note that it is not feasible to set the beamforming vector of the XL-MIMO array as in {\bl\eqref{Eq: Section2_1_8}}, due to the constant amplitude of an analog phase shifter. It has been demonstrated that an Airy beam can be constructed by an XL-MIMO array in engineering following the cubic phase method \cite{Works4}. According to {\bl\eqref{Eq: Section2_1_7}}, the phase of the analog phase shifter that connects to the $\left(m ,n\right)$-th antenna (i.e., the antenna in the $m$-th row and the $n$-th column) is set as $\frac{x_0^3}{3} x_n^3 + \frac{y_0^3}{3} y_m^3$, for $m = 1,\ldots,M$ and $n = 1,\ldots,N$, where $y_0 \ne 0$ is a transverse scale that controls the curvature of an Airy beam in the $y-z$ plane. When the BS transmits a pilot that is equal to 1, the signal received on the position $\left(x,y,z\right)$, for any $z > 0$, can be calculated by the Rayleigh-Sommerfeld diffraction integral function as \cite{Works3}
\begin{equation}\label{Eq: Section2_2_1}
\begin{aligned}
s \left(x,y,z\right) = &\sqrt{\frac{P}{MN}} \sum\limits_{n = 1}^{N} \sum\limits_{m = 1}^{M} \exp\left[j \frac{x_0^3}{3}x_n^3 + j \frac{y_0^3}{3}y_m^3 \right] \\ \times & {\rm Green} \left(x,y,z,x_n,y_m\right),
\end{aligned}
\end{equation}
where $P$ is the transmit power; $x_n$ and $y_m$ denote the $x$-axis and $y$-axis positions of the $\left(m, n\right)$-th antenna, such that $x_n = \left(n - \frac{N+1}{2}\right)d$ and $y_m = \left(m - \frac{M+1}{2}\right)d$, respectively; ${\rm Green} \left(x,y,z,x_n,y_m\right)$ is the Green's function, such that
\begin{equation}\label{Eq: Section2_2_2}
\begin{aligned}
&{\rm Green} \left(x,y,z,x_n,y_m\right) \\= &\frac{\lambda}{4 \pi r} \exp \left[ jk\sqrt{\left(x-x_n\right)^2 + \left(y-y_m\right)^2 + z^2} \right] \\ \mathop \approx \limits^{\left(\rm a\right)} & \frac{\lambda e^{jkz}}{4 \pi r} \underbrace{\exp \left[ j \frac{k \left(x-x_n\right)^2}{2z} \right]}_{n-{\rm only}} \underbrace{\exp \left[ j \frac{k \left(y-y_m\right)^2}{2z} \right]}_{m-{\rm only}},
\end{aligned}
\end{equation}
where $r = \sqrt{x^2 + y^2 + z^2}$ represents the distance between the center of the UPA and the position $\left(x,y,z\right)$, and (a) is derived by $\sqrt{1 + t} = 1 + \frac{1}{2}t + {\cal O}\left(t^2\right)$ (i.e., the paraxial propagation condition, where the ${\cal O} \left(z^{-3}\right)$ item is omitted). Substituting {\bl\eqref{Eq: Section2_2_2}} into {\bl\eqref{Eq: Section2_2_1}}, we can get
\begin{equation}\label{Eq: Section2_2_3}
\begin{aligned}
s \left(x,y,z\right) \approx \frac{ \sqrt{P} \lambda e^{jkz}}{ 4\pi \sqrt{M N} r} {\psi}_{{\rm A}, x} \left(x,z\right) {\psi}_{{\rm A}, y} \left(y,z\right),
\end{aligned}
\end{equation}
where ${\psi}_{{\rm A}, x} \left(x,z\right)$ is the EF in the $x-z$ plane, such that
\begin{equation}\label{Eq: Section2_2_4}
{\psi}_{{\rm A}, x} \left(x,z\right) = \sum\limits_{n = 1}^{N} {\exp \left[j \left(\frac{x_0^3}{3}x_n^3 + \frac{k\left(x - x_n\right)^2}{2z} \right) \right]},
\end{equation}
while ${\psi}_{{\rm A}, y} \left(y,z\right)$ denotes the EF in the $y-z$ plane, such that
\begin{equation}\label{Eq: Section2_2_5}
{\psi}_{{\rm A}, y} \left(y,z\right) = \sum\limits_{m = 1}^{M} {\exp \left[j \left(\frac{y_0^3}{3}y_m^3 + \frac{k\left(y - y_m\right)^2}{2z} \right) \right]}.
\end{equation}
Note that similar to {\bl\eqref{Eq: Section2_1_2}}, the EF has also been decoupled from the $x$ and the $y$ directions. Therefore, only the EF in the $x-z$ plane will then be analyzed, while a similar solution can be obtained w.r.t. the EF in the $y-z$ plane. The subscript $y$ is omitted in the following sections.\footnote{Note that the spatial-domain parameters $x$ and $z$ can also converted to the polar-domain domain parameters $r$ and $\theta$, i.e., $x = r \sin \left(\theta \right)$ and $z = r \cos \left(\theta \right)$, where $r = \sqrt{x^2 + z^2}$ is the distance between the position $\left(x, z\right)$ and the origin, while $\theta$ denotes the azimuth angle \cite{XLMIMOSystems4}. Moreover, the received signal is typically rewritten as $s\left(x, z\right) = \sqrt{P} {\bf w}^{\rm H} {\bf h}$ in some literature (in the $x-z$ plane and without noise), where ${\bf h} \in \mathbb{C}^{N \times 1}$ denotes the channel, while ${\bf w} \in \mathbb{C}^{N \times 1}$ represents the beamforming vector is this paper.}

Nevertheless, this simplified implementation for Airy beams in XL-MIMO systems (i.e., the cubic phase method) will lead to fundamental changes in beam properties compared to {\bl\cite{UnifiedTheory}}. In addition, the finite and discrete arrays are equivalent to the sequential execution of a ``truncation'' and ``sampling'' operations on the infinite and continuous $x-z$ plane, which imposes constraints on minimum array aperture and maximum antenna spacing for a distortion-free main lobe trajectory. Consequently, it is essential to establish an analytical theoretical framework of Airy beams in XL-MIMO systems.

\section{Analytical Theoretical Framework of\\Airy Beams in XL-MIMO Systems}\label{Sec: Theoretical Framework}

In this section, actual main lobe trajectories of Airy beams in XL-MIMO systems, as well as constraints on array aperture and antenna spacing, are systematically analyzed.

\subsection{Actual Beam Properties in XL-MIMO Systems}

It is important to note that, when the XL-MIMO system is deployed with a significantly larger array aperture and a much smaller antenna spacing, i.e., $L \to + \infty$ and $d \to 0$, the actual EF can be approximated as an infinite EF, such that
\begin{equation}\label{Eq: Section3_1_1}
\psi_{\rm A} \left(x,z\right) |_{L \to + \infty}^{d \to 0} \to \psi_{\rm I} \left(x,z\right),
\end{equation}
for any $z > 0$. The infinite EF is given by
\begin{equation}\label{Eq: Section3_1_2}
{\psi}_{\rm I} \left(x, z\right) = \int_{-\infty}^{+\infty} \psi \left(x', 0 \right) {\exp \left[j \frac{k\left(x - x'\right)^2}{2z} \right]} dx'.
\end{equation}
Note that $\psi \left( x', 0 \right) = \exp \left[j \omega \left( x' \right) \right]$ denotes the beamforming vector of the XL-MIMO array, since $z = 0$, where $\omega \left( x' \right) \in \left[0, 2\pi\right)$ is the phase set by an analog phase shifter on the position $\left(x', 0\right)$ along the $x$-axis.

\subsubsection{Cubic Phase Method}

When the cubic phase method is adopted to construct an Airy beam \cite{Works3}, i.e., $\omega \left(x'\right) = \frac{x_0^3}{3}x'^3$, the beamforming vector can be denoted as
\begin{equation}\label{Eq: Section3_1_3}
{\psi} \left(x',0\right) = \exp\left[j\frac{x_0^3}{3}x'^3\right].
\end{equation}

\textbf{\textit{Theorem 1:}} The EF on the position $\left(x, z\right)$ can be given by
\begin{equation}\label{Eq: Section3_1_4}
\begin{aligned}
\psi_{\rm I} \left(x, z \right) = &\frac{1}{x_0} {\rm Ai} \left( -\frac{k^2}{4z^2 x_0^4} - \frac{kx}{zx_0} \right) \\ \times & \exp \left[j \left(\frac{k^3}{12z^3x_0^6} + \frac{k^2 x}{2 z^2 x_0^3} + \frac{kx^2}{2z} \right) \right],
\end{aligned}
\end{equation}
for any $z > 0$, while the main lobe trajectory is expressed as
\begin{equation}\label{Eq: Section3_1_5}
x \left(z \right) = \underbrace {-\frac{k}{4 x_0^3} \frac{1}{z}}_{{\rm Part}\: 1} 
+ \underbrace {\left(- \frac{\mu_0x_0}{k}\right) z}_{{\rm Part}\: 2},
\end{equation}
which is a hyperbolic trajectory (i.e., Part 1) plus a straight-line trajectory (i.e., Part 2) along $z$.

\textbf{\textit{Proof:}} See Appendix A.

\textbf{\textit{Proposition 1:}} When $0 < z \ll z_{\rm c}$, {\bl\eqref{Eq: Section3_1_5}} is approximated to be a hyperbolic trajectory, i.e., $x\left(z\right) \approx -\frac{k}{4 x_0^3} \frac{1}{z}$, while when $z$ becomes larger and satisfies $z \gg z_{\rm c}$, {\bl\eqref{Eq: Section3_1_5}} is approximated to be a straight-line trajectory, i.e., $x\left(z\right) \approx - \frac{\mu_0x_0}{k} z$, where $z_{\rm c}$ is calculated by
\begin{equation}\label{Eq: Section3_1_6}
-\frac{k}{4x_0^3} \frac{1}{z_{\rm c}} = - \frac{\mu_0 x_0}{k} z_{\rm c},
\end{equation}
such that
\begin{equation}\label{Eq: Section3_1_7}
z_{\rm c} = \frac{k}{2 x_0^2 \sqrt{\left|\mu_0 \right|}}.
\end{equation}
Therefore, it can be concluded that
\begin{equation}\label{Eq: Section3_1_8}
\left\{ {\begin{array}{*{20}{l}}
{x\left(z\right) \approx -\frac{k}{4x_0^3} \frac{1}{z},\:\:\:\:\:\: \:\:\:\:\:\: \:\:\:\:\:\: \:\, 0 < z \ll z_{\rm c}}\\
{x\left(z\right) = -\frac{k}{4x_0^3} \frac{1}{z} - \frac{\mu_0 x_0}{k} z,\:\:\: {\rm others}}\\
{x\left(z\right) \approx - \frac{\mu_0 x_0}{k} z,\:\:\:\:\:\: \:\:\:\:\:\: \:\:\:\:\,\,\, z \gg z_{\rm c}}
\end{array}} \right..
\end{equation}
A typical case is given, where $x_0 = 1$: In an upper-6 GHz (U6G) frequency band that satisfies $f_{\rm c} = 7$ GHz, $z_{\rm c} \approx 72.6247$ m, while in a millimeter wave (mmWave) frequency band that satisfies $f_{\rm c} = 30$ GHz, $z_{\rm c} \approx 311.2487$ m.

\textbf{\textit{Proposition 2:}} Note that the transverse trajectory velocity of the Airy beam in {\bl\eqref{Eq: Section3_1_4}} can be calculated as
\begin{equation}\label{Eq: Section3_1_9}
\frac{\partial x\left(z\right)}{\partial z} = \frac{k}{4x_0^3} \frac{1}{z^2} - \frac{\mu_0 x_0}{k}.
\end{equation}
Therefore, it can be concluded that $x\left(z\right)$ is a monotonically increasing function along $z$ when $x_0 > 0$ (i.e., from $-\infty$ to $+\infty$), while it is a monotonically decreasing function when $x_0 < 0$ (i.e., from $+\infty$ to $-\infty$), for $z > 0$.

\subsubsection{Cubic Phase Plus Focusing Phase Method}

In wireless engineering, the beamforming vector is typically set as a composite of cubic and focusing phases, enabling both curved trajectories and beam focusing \cite{Works5}. The focusing position is indicated by $\left(x_{\rm F}, z_{\rm F}\right)$, where $z_{\rm F} > 0$, while $\omega \left(x'\right) = \frac{x_0^3}{3}x'^3 - \frac{k \left(x_{\rm F} - x'\right)^2}{2z_{\rm F}}$. Then, the EF is expressed as
\begin{equation}\label{Eq: Section3_1_10}
{\psi}_{\rm I} \left(x, z\right) = \int_{-\infty}^{+\infty} \psi \left(x', 0, x_{\rm F}, z_{\rm F} \right) {\exp \left[j \frac{k\left(x - x'\right)^2}{2z} \right]} dx',
\end{equation}
for any $z > 0$, where the beamforming vector is given by
\begin{equation}\label{Eq: Section3_1_11}
\psi \left(x', 0, x_{\rm F}, z_{\rm F} \right) = \exp\left[j \left(\frac{x_0^3}{3}x'^3 - \frac{k\left(x_{\rm F} - x'\right)^2}{2z_{\rm F}}\right)\right].
\end{equation}

\begin{figure*}[ht]
\centering
\begin{equation}\label{Eq: Section3_1_12}
\psi_{\rm I} \left(x, z\right) = \frac{1}{x_0} {\rm Ai} \left( - \frac{k^2 \left(z_{\rm F} - z \right)^2}{4 z^2 z_{\rm F}^2 x_0^4} -\frac{k\left(xz_{\rm F} - x_{\rm F}z \right)}{zz_{\rm F} x_0} \right) \exp {\left[j \frac{k^3 \left(z_{\rm F} - z\right)^3}{12z^3 z_{\rm F}^3 x_0^6} + j\frac{k^2 \left(xz_{\rm F} - x_{\rm F}z \right) \left(z_{\rm F} - z \right)}{2z^2 z_{\rm F}^2 x_0^3} + j\frac{k \left(x^2z_{\rm F} - x_{\rm F}^2 z\right)}{2z z_{\rm F}} \right]}
\end{equation}
\hrule
\end{figure*}

\textbf{\textit{Theorem 2:}} In the cubic phase plus focusing phase method, the EF on the position $\left(x, z\right)$ can be expressed as {\bl\eqref{Eq: Section3_1_12}}. Then, similar to \cite{Works5}, the main lobe trajectory is calculated as
\begin{equation}\label{Eq: Section3_1_13}
x\left(z\right) = -\frac{k}{4 x_0^3} \frac{1}{z} - \frac{\mu_0x_0}{k} z + \underbrace {\left( \frac{x_{\rm F}}{z_{\rm F}} - \frac{k}{4z_{\rm F}^2 x_0^3} \right) z + \frac{k}{2 z_{\rm F} x_0^3}}_{{\rm Part}3}.
\end{equation}
Compared to {\bl\eqref{Eq: Section3_1_5}}, the focusing phase results the trajectory in a straight-line offset along $z > 0$ and a transverse offset as in Part 3, with a slope of $\frac{x_{\rm F}}{z_{\rm F}} - \frac{k}{4z_{\rm F}^2 x_0^3}$ and an intercept of $\frac{k}{2 z_{\rm F} x_0^3}$.

\textbf{\textit{Proof:}} See Appendix B.

Note that the transverse trajectory velocity of the Airy beam in {\bl\eqref{Eq: Section3_1_12}} can be calculated as
\begin{equation}\label{Eq: Section3_1_14}
\frac{\partial x\left(z\right)}{\partial z} = \underbrace {\frac{k}{4x_0^3} \frac{1}{z^2}}_{{\rm Part}\:4} + \underbrace {\frac{x_{\rm F}}{z_{\rm F}} - \frac{k}{4 z_{\rm F}^2 x_0^3} - \frac{\mu_0 x_0}{k}}_{{\rm Part}\:5},
\end{equation}
for $z > 0$, while the transverse trajectory acceleration is
\begin{equation}\label{Eq: Section3_1_15}
\frac{\partial^2 x\left(z\right)}{\partial z^2} = - \frac{k}{2 x_0^3} \frac{1}{z^3}.
\end{equation}

\textbf{\textit{Proposition 3:}} The main lobe trajectory as in {\bl\eqref{Eq: Section3_1_13}} retains the non-diffraction property but no longer exhibits the self-acceleration property. It can be divided into four types:

Type 1: The main lobe trajectory is a concave function along $z > 0$ (that from $-\infty$ increases to $x_{\rm ext}$ and then decreases to $-\infty$) when $x_0 > 0$ and
\begin{equation}\label{Eq: Section3_1_16}
x_{\rm F} < \frac{k}{4 x_0^3}\frac{1}{z_{\rm F}} + \frac{\mu_0 x_0}{k}z_{\rm F},
\end{equation}
with the extreme maximum value is located at
\begin{equation}\label{Eq: Section3_1_17}
z_{\rm ext} = \frac{kz_{\rm F}}{\sqrt{k^2 - 4kx_{\rm F} z_{\rm F} x_0^3 + 4 \mu_0 z_{\rm F}^2 x_0^4}},
\end{equation}
while the extreme maximum value is
\begin{equation}\label{Eq: Section3_1_18}
x_{\rm ext} = x\left(z_{\rm ext}\right) = \frac{k}{2 x_0^3} \left(\frac{1}{z_{\rm F}} - \frac{1}{z_{\rm ext}}\right).
\end{equation}

Type 2: The main lobe trajectory is a monotonically increasing function along $z > 0$ when $x_0 > 0$ and $x_{\rm F} \ge \frac{k}{4 x_0^3} \frac{1}{z_{\rm F}} + \frac{\mu_0 x_0}{k}z_{\rm F}$ (that from $-\infty$ to $+ \infty$).

Type 3: Similar to Type 1, the main lobe trajectory is a convex function along $z > 0$ (that from $+\infty$ decreases to $x_{\rm ext}$ and then increases to $+\infty$) when $x_0 < 0$ and $x_{\rm F} > \frac{k}{4 x_0^3}\frac{1}{z_{\rm F}} + \frac{\mu_0 x_0}{k} z_{\rm F}$. The extreme minimum value is located at $z = z_{\rm ext}$, while the minimum extreme value is $x_{\rm ext}$.

Type 4: Similar to Type 2, the main lobe trajectory is a monotonically decreasing function along $z > 0$ when $x_0 < 0$ and $x_{\rm F} \le \frac{k}{4 x_0^3} \frac{1}{z_{\rm F}} + \frac{\mu_0 x_0}{k} z_{\rm F}$ (that from $+ \infty$ to $- \infty$).

\textbf{\textit{Proof:}} See Appendix C.

\textbf{\textit{Proposition 4:}} When $z_{\rm F} \ll z_{\rm c}$, the main lobe trajectory passes through the focusing position $\left(x_{\rm F}, z_{\rm F} \right)$, such that
\begin{equation}\label{Eq: Section3_1_19}
x\left(z_{\rm F}\right) = x_{\rm F} - \frac{\mu_0 x_0 c}{2\pi f_{\rm c}} z_{\rm F} \approx x_{\rm F},
\end{equation}
where $-\frac{\mu_0 x_0 c}{2\pi f_{\rm c}} z_{\rm F} \to 0$ when $z_{\rm F} \ll z_{\rm c}$. In addition, the error term $-\frac{\mu_0 x_0 c}{2\pi f_{\rm c}} z_{\rm F}$ further reduces with an increase in $f_{\rm c}$.

In particular, the following special case is considered: $x_{\rm F} = 0$ and $0 < z \ll z_{\rm c}$. Then, the main lobe trajectory is given by
\begin{equation}\label{Eq: Section3_1_20}
x\left(z\right) \approx -\frac{k}{4 x_0^3} \frac{1}{z} - \frac{k}{4z_{\rm F}^2 x_0^3}z + \frac{k}{2 z_{\rm F} x_0^3},
\end{equation}
and
\begin{equation}\label{Eq: Section3_1_21}
\frac{\partial x\left(z\right)}{\partial z} \approx \frac{k}{4x_0^3} \frac{1}{z^2} - \frac{k}{4 z_{\rm F}^2 x_0^3}.
\end{equation}
Note that the formula {\bl\eqref{Eq: Section3_1_20}} must belong to Type 1 or Type 3, while the extreme maximum/minimum value is located at the focusing position $\left(x_{\rm ext}, z_{\rm ext}\right) = \left(x_{\rm F}, z_{\rm F}\right)$.

\begin{figure*}
  \centering
  \includegraphics[scale=0.34]{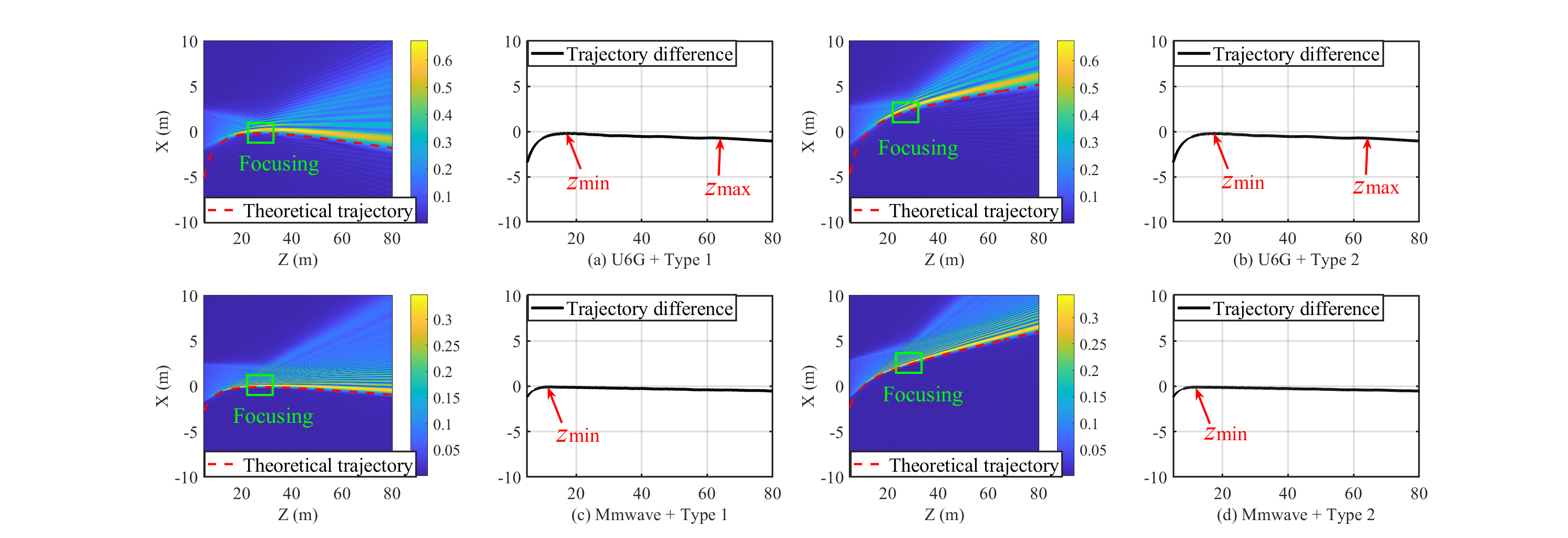}
  \caption{Differences between the theoretical and actual main lobe trajectories, where $d = \frac{\lambda}{2}$. A U6G XL-MIMO system: $f_{\rm c} = 7$ GHz, $N = 256$, and $x_0 = 1$; a mmWave XL-MIMO system: $f_{\rm c} = 30$ GHz, $N = 1024$, and $x_0 = 2$. Type 1: $\left(x_{\rm F}, z_{\rm F}\right) = \left(0, 30\right)$ m; Type 2: $\left(x_{\rm F}, z_{\rm F}\right) = \left(2.6147, 29.8858\right)$ m.} \label{Fig: simulation1}
\end{figure*}

The difference between the theoretical main lobe trajectory, $x\left(z\right)$, and the actual main lobe trajectory, $\tilde x\left(z\right)$, is evaluated against $z$, which is defined as $\Delta x \left(z\right) = x\left(z\right) - \tilde x\left(z\right)$, where $z > 0$. As shown in {\bl Fig.~\ref{Fig: simulation1}}, across the U6G frequency band and mmWave frequency band and for various main lobe trajectory types, the theoretical main lobe trajectory of XL-MIMO systems closely matches the actual main lobe trajectory, with negligible deviation$\Delta x \left(z\right)$.

\subsection{Array Constraints for Truncation and Sampling}

In practical XL-MIMO systems, the array aperture is finite with a length $L$ (that from $-\frac{L}{2}$ to $+\frac{L}{2}$), and discrete with antenna spacing $d$. This is tantamount to performing a ``truncation'' and ``sampling'' operations on $\psi \left(x',0\right)$, i.e.,
\begin{equation}\label{Eq: Section3_2_1}
\psi \left(x_n, 0, x_{\rm F}, z_{\rm F} \right) = \psi \left(x', 0, x_{\rm F}, z_{\rm F} \right) |_{x' = x_n},
\end{equation}
for $n = 1,\ldots,N$. The error between the infinite EF $\psi_{\rm I} \left(x,z\right)$ and the actual EF $\psi_{\rm A} \left(x,z\right)$ on the position $\left(x,z\right)$ can be expressed as
\begin{equation}\label{Eq: Section3_2_2}
\begin{aligned}
&\epsilon \left(x, z \right) = \psi_{\rm I} \left(x,z\right) - \psi_{\rm A} \left(x,z\right) \\ = & \underbrace {\psi_{\rm I} \left(x,z\right) - \psi_{\rm M} \left(x,z\right)}_{\epsilon_1 \left(x, z \right)} + \underbrace {\psi_{\rm M} \left(x,z\right) - \psi_{\rm A} \left(x,z\right)}_{\epsilon_2 \left(x, z \right)},
\end{aligned}
\end{equation}
for any $z > 0$, where $\psi_{\rm I}\left(x, z \right)$ is denoted as {\bl\eqref{Eq: Section3_1_10}}, while $\psi_{\rm A} \left(x, z\right)$ and the finite EF $\psi_{\rm M} \left(x,z\right)$ on the position $\left(x,z\right)$ are respectively defined as
\begin{equation}\label{Eq: Section3_2_3}
\psi_{\rm A} \left(x, z \right) = d \sum \limits_{n = 1}^N {\psi \left(x_n,0,x_{\rm F}, z_{\rm F}\right) \exp\left[j \frac{k\left(x - x_n\right)^2}{2z}\right]},
\end{equation}
and
\begin{equation}\label{Eq: Section3_2_4}
\psi_{\rm M} \left(x, z \right) = \int_{-\frac{L}{2}}^{+\frac{L}{2}} {\psi \left(x',0,x_{\rm F}, z_{\rm F}\right) \exp\left[j \frac{k\left(x - x'\right)^2}{2z}\right]} d x'.
\end{equation}
Moreover, $\epsilon_1 \left(x, z \right)$ and $\epsilon_2 \left(x, z \right)$ represent the truncation error and the sampling error, respectively.

\subsubsection{Truncation Constraints}

To ensure a distortion-free main lobe trajectory under truncation, the minimum required array aperture must be determined.

\textbf{\textit{Lemma 1}}: The primary contribution of an oscillatory integral $\int_{-\infty}^{+ \infty} {\exp \left[j \phi \left(x'\right) \right]} dx'$ on the position $\left(x,z\right)$ is attributable to the stationary points within the array aperture \cite{References4}, for any $z > 0$. The stationary points can be calculated by
\begin{equation}\label{Eq: Section3_2_5}
\phi' \left(x'\right) = 0,
\end{equation}
where $\phi' \left(x'\right)$ represents the first-order partial derivative of $x'$. In particular, according to the MSP method, the main lobe trajectory of an Airy beam is determined by the caustics (where the two stationary points of an Airy beam merge into the same point, i.e., a second-order stationary point) \cite{References5}, while the caustics satisfies both {\bl\eqref{Eq: Section3_2_5}} and
\begin{equation}\label{Eq: Section3_2_6}
\phi'' \left(x'\right) = 0,
\end{equation}
where $\phi'' \left(x'\right)$ is the second-order partial derivative of $x'$. \qed

\textbf{\textit{Theorem 3:}} Given the minimum and maximum $z$-axis distances, $z_{\rm min} >0$ and $z_{\rm max} > 0$ (i. e., the main lobe trajectory remains unchanged after the truncation operation for $z_{\rm min} \le z \le z_{\rm max}$), it is necessary that $L$ satisfies
\begin{equation}\label{Eq: Section3_2_7}
L \ge \max \left\{ \frac{1}{z_{\rm min}} - \frac{1}{z_{\rm F}}, \frac{1}{z_{\rm F}} - \frac{1}{z_{\rm max}} \right\} \frac{k}{\left|x_0^3\right|}  + 2 \left|3 x_0^{-3}\right|^{\frac{1}{3}}.
\end{equation}
Thus, given a specified $L > 2 \left|3 x_0^{-3}\right|^{\frac{1}{3}}$, the minimum and maximum propagation distances (between which the main lobe trajectory is distortion-free) are
\begin{equation}\label{Eq: Section3_2_8}
z_{\min} = \frac{kz_{\rm F}}{\left(L - 2\left|3 x_0^{-3}\right|^{\frac{1}{3}} \right)z_{\rm F} \left|x_0^3 \right| + k},
\end{equation}
and
\begin{equation}\label{Eq: Section3_2_9}
z_{\max} = \frac{kz_{\rm F}}{k - \left(L - 2\left|3 x_0^{-3}\right|^{\frac{1}{3}} \right) z_{\rm F} \left|x_0^3 \right|},
\end{equation}
while $z_{\rm max} \to +\infty$ when $k \le \left(L - 2\left|3 x_0^{-3}\right|^{\frac{1}{3}} \right) z_{\rm F} \left|x_0^3 \right|$. It can be concluded that the main lobe trajectory is always distortion-free close to the focusing point $\left(x_{\rm F}, z_{\rm F}\right)$.

\textbf{\textit{Proof:}} See Appendix D.

\textbf{\textit{Lemma 2:}} Based on the oscillation integral theorem, when $\phi \left(x'\right)$ is a monotonic function and $\phi' \left(x'\right) \ne 0$, for $a \le x' \le b$, we can get \cite{References6}
\begin{equation}\label{Eq: Section3_2_10}
\left|\int_{a}^{b} {\exp\left[j \phi \left(x' \right) \right]} dx'\right| \le \frac{2}{\min_{a \le x' \le b} \left| \phi'\left(x'\right)\right|}.
\end{equation}
\qed

\textbf{\textit{Proposition 5:}} Note that $\epsilon_1\left(x, z \right)$ can be expressed as
\begin{equation}\label{Eq: Section3_2_11}
\epsilon_1\left(x, z \right) = \int_{x' > \left|\frac{L}{2}\right|} {\psi \left(x',0, x_{\rm F}, z_{\rm F} \right) \exp \left[j \frac{k\left(x - x'\right)^2}{2z}\right]} dx'.
\end{equation}
for $z > 0$. Thus, on the main lobe trajectory, the truncation error $\epsilon_1 \left(x, z \right)$ is proportional to the inverse of $L^2$, i.e.,
\begin{equation}\label{Eq: Section3_2_12}
\left|\epsilon_1 \left(x, z \right) \right| \sim {\cal O} \left(L^{-2}\right).
\end{equation}

\textbf{\textit{Proof:}} See Appendix E.

\subsubsection{Sampling Constraints}

Subsequent to the sampling operation, the continuous plane degenerates into a discrete array. Then, the equivalent stationary points along a discrete array will be calculated by a first-order symmetric difference equation \cite{References7}, which is defined as
\begin{equation}\label{Eq: Section3_2_13}
D_{\rm d}^{\left(1\right)} \phi \left(x'\right) = \frac{\phi \left(x' + d\right) - \phi \left(x' - d\right)}{2d} = 0,
\end{equation}
while the equivalent caustics is calculated by a second-order symmetric difference equation, such that
\begin{equation}\label{Eq: Section3_2_14}
D_{\rm d}^{\left(2\right)} \phi \left(x'\right) = \frac{D_{\rm d}^{\left(1\right)} \phi \left(x' + d\right) - D_{\rm d}^{\left(1\right)} \phi \left(x' - d\right)}{2d} = 0.
\end{equation}

\textbf{\textit{Theorem 4:}} Subsequent to the execution of the sampling operation, a trajectory offset will be generated compared to {\bl\eqref{Eq: Section3_1_13}}, which can be expressed as
\begin{equation}\label{Eq: Section3_2_15}
\Delta x\left(z\right) = \frac{d^2 x_0^3}{3k}z,
\end{equation}
for $z > 0$, while it can typically be ignored due to the much smaller item ${\cal O} \left(d^2\right)$ in practical XL-MIMO systems.

\textbf{\textit{Proof:}} See Appendix F.

\textbf{\textit{Proposition 6:}} The antenna spacing should satisfy
\begin{equation}\label{Eq: Section3_2_16}
d \le \min \left\{ \frac{4 \pi z_{\rm F}}{\left|z_{\rm F} x_0^3 L^2 \pm 2kL + 4kx_{\rm F}\right|}, \frac{\lambda}{2} \right\},
\end{equation}
where $d \le \frac{\lambda}{2}$ is a classic constraint in an XL-MIMO system.

\textbf{\textit{Proof:}} See Appendix G.

\textbf{\textit{Lemma 3:}} According to the Euler-Maclaurin error bound, the sampling error upper bound with an array aperture $L$ and sampling spacing $d$ is given by \cite{References8}
\begin{equation}\label{Eq: Section3_2_17}
\left| \int_{-\frac{L}{2}}^{+\frac{L}{2}} {f \left(x' \right)} dx' - d \sum\limits_{n = 1}^N f\left(x_n \right) \right| \le \max_{-\frac{L}{2} \le x' \le \frac{L}{2}} \left| \frac{\partial f\left(x'\right)}{\partial x'}  \right| \frac{dL}{2},
\end{equation}
where $f \left(x'/x_n\right)$ is a function of $x'/x_n$, for $n = 1,\ldots,N$. \qed

\begin{figure}
  \centering
  \includegraphics[scale=0.3]{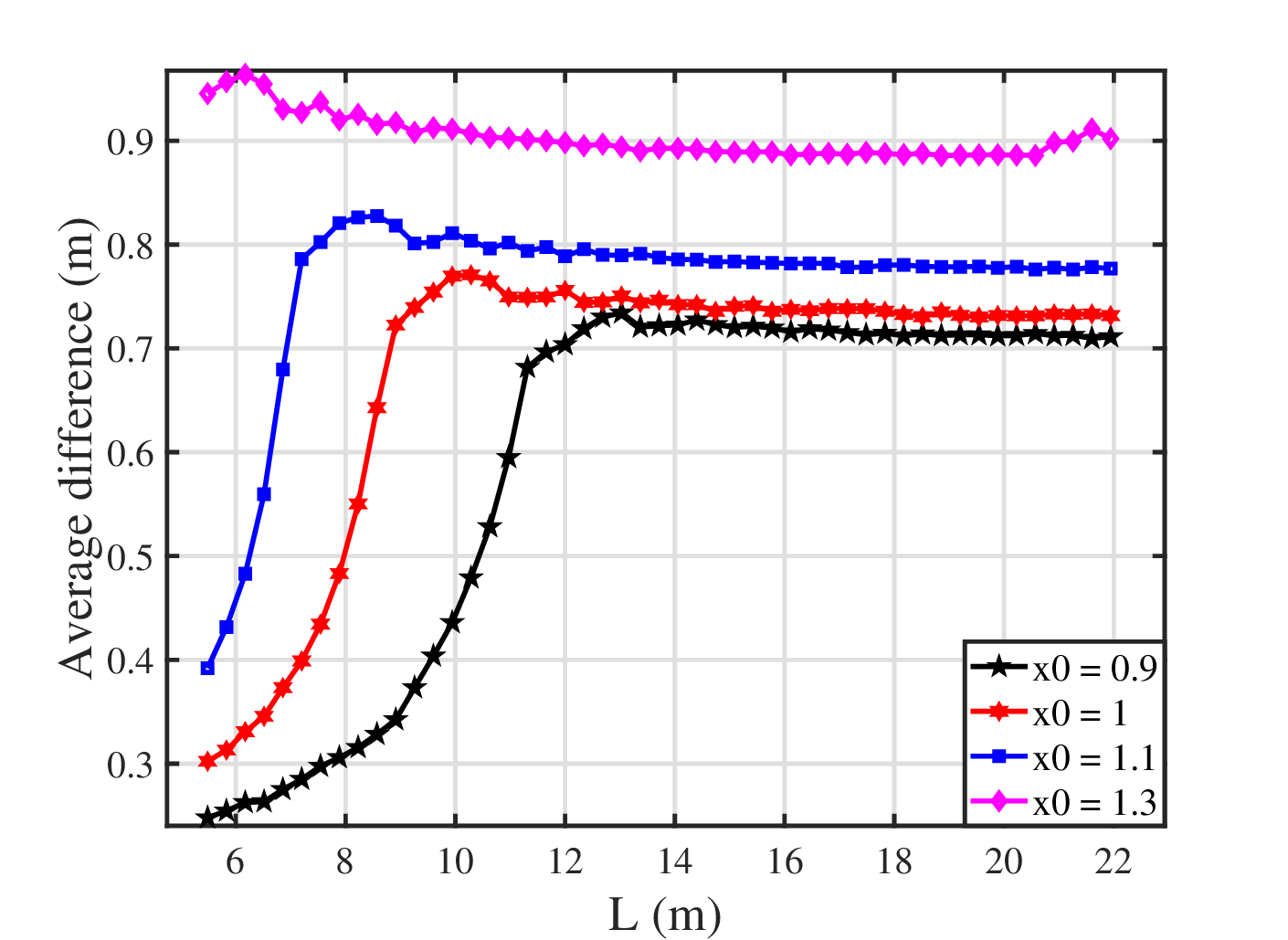}
  \caption{$\Delta \bar x$ against $L$, where $f_{\rm c} = 7$ GHz and $d = \frac{\lambda}{2}$.} \label{Fig: simulation2}
\end{figure}

\textbf{\textit{Proposition 7:}} According to \textbf{\textit{Lemma 3}}, the sampling error $\epsilon_2 \left(x, z\right)$ on the main lobe trajectory satisfies
\begin{equation}\label{Eq: Section3_2_18}
\left| \epsilon_2 \left(x,z\right) \right| \le \max_{-\frac{L}{2} \le x' \le \frac{L}{2}} \left| \frac{\partial \psi \left(x', 0, x_{\rm F}, z_{\rm F} \right)}{\partial x'}  \right| \frac{dL}{2},
\end{equation}
for $z > 0$, where
\begin{equation}\label{Eq: Section3_2_19}
\left| \frac{\partial \psi \left(x', 0, x_{\rm F}, z_{\rm F} \right)}{\partial x'} \right| = \left| j \phi' \left(x' \right) \psi \left(x', 0, x_{\rm F}, z_{\rm F} \right)\right| = \left| \phi' \left(x' \right) \right|,
\end{equation}
since $\left|j \psi \left(x', 0, x_{\rm F}, z_{\rm F} \right) \right| = 1$. Therefore, we can get
\begin{equation}\label{Eq: Section3_2_20}
\left| \epsilon_2 \left(x,z\right) \right| \le \max_{-\frac{L}{2} \le x' \le \frac{L}{2}} \left| \phi' \left(x'\right) \right| \frac{dL}{2},
\end{equation}
where $\phi' \left(x'\right)$ is defined as in {\bl\eqref{Eq: SectionD_1}}. Subsequently, according to {\bl\eqref{Eq: SectionE_2}} and {\bl\eqref{Eq: SectionE_3}}, we have
\begin{equation}\label{Eq: Section3_2_21}
\max_{-\frac{L}{2} \le x' \le \frac{L}{2}} \left| \phi' \left(x' \right) \right| = \max \left|\phi' \left(\pm \frac{L}{2}\right)  \right| \sim {\cal O} \left(L^2\right).
\end{equation}
Thus, the sampling error $\epsilon_2 \left(x, z\right)$ is proportional to $d L^3$, i.e.,
\begin{equation}\label{Eq: Section3_2_22}
\left| \epsilon_2 \left(x,z\right) \right| \sim {\cal O} \left(d L^3\right).
\end{equation}

It can be concluded that, to ensure a distortion-free main lobe trajectory and minimize the error, the truncation operation requires a larger $L$ as in {\bl\eqref{Eq: Section3_2_12}}, while the sampling operation requires a smaller $d$ as $L$ increases as in {\bl\eqref{Eq: Section3_2_16}} and {\bl\eqref{Eq: Section3_2_22}}.

\begin{figure}
  \centering
  \includegraphics[scale=0.3]{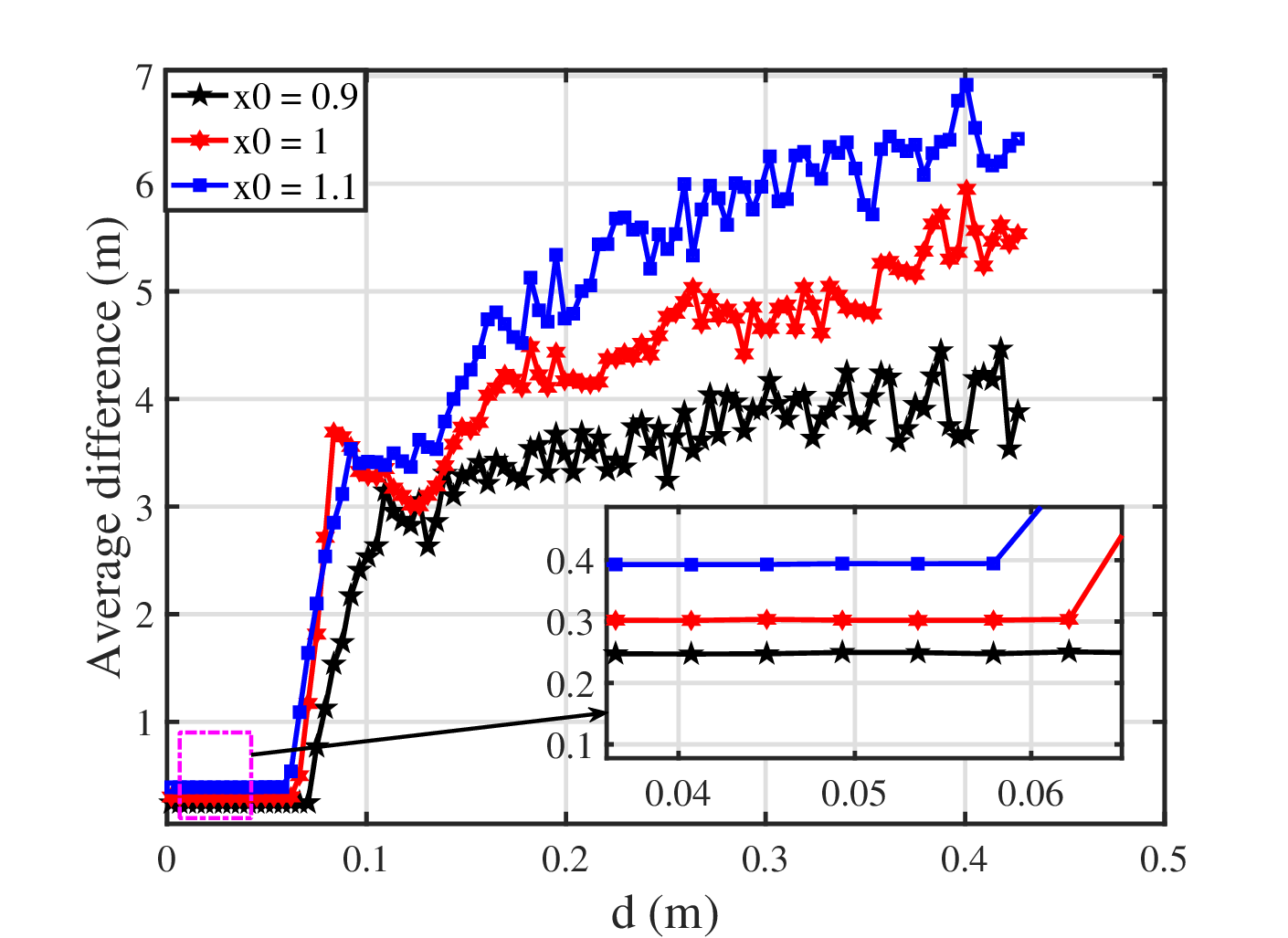}
  \caption{$\Delta \bar x$ against $d$, where $f_{\rm c} = 7$ GHz and $L = 5.4857$ m.} \label{Fig: simulation3}
\end{figure}

As shown in {\bl Fig.~\ref{Fig: simulation1}}, the theoretical $z_{\rm min}$ and $z_{\rm max}$ can effectively determine the distortion-free main lobe trajectory range after the truncation and sampling operations. The average differences between the theoretical and actual main lobe trajectories against $L$ and $d$ are then evaluated, respectively, where $\left(x_{\rm F}, z_{\rm F}\right) = \left(1.7431, 19.9239\right)$ m. The average difference, $\Delta \bar x$, is defined as
\begin{equation}\label{Eq: Section5_1}
\Delta \bar x = \frac{1}{z_{\rm max} - z_{\rm min}} \int_{z_{\rm min}}^{z_{\rm max}} {\left| x\left(z\right) - \tilde x\left(z\right)\right|} dz.
\end{equation}
As shown in {\bl Fig.~\ref{Fig: simulation2}}, $\Delta \bar x$ exhibits a non-monotonic behavior as $L$ increases. Specifically, $\Delta \bar x$ first increases rapidly, which can be attributed to the presence of sampling error scaling with $L^3$ as in {\bl\eqref{Eq: Section3_2_22}}. However, when $L$ becomes sufficiently large, the distortion-free main lobe trajectory range $\left[x_{\min}, x_{\max}\right]$ shrinks, as indicated by {\bl\eqref{Eq: Section3_2_8}} and {\bl\eqref{Eq: Section3_2_9}}, both of which gradually converge toward the focusing point. As a result, $\Delta \bar x$ subsequently decreases and eventually approaches a stable value. Moreover, $\Delta \bar x$ increases with $\left| x_0 \right|$, due to the main lobe trajectory offset scaling with $\left|x_0^3\right|$ as in {\bl\eqref{Eq: Section3_2_15}}, while a larger $\left|x_0\right|$ also leads to an earlier convergence of $\Delta \bar{x}$. As shown in {\bl Fig.~\ref{Fig: simulation3}}, $\Delta \bar x$ will also increase with both $d$ and $\left|x_0\right|$, due to the sampling error as in {\bl\eqref{Eq: Section3_2_22}}, which scales with $d$, and the main lobe trajectory offset as in {\bl\eqref{Eq: Section3_2_15}}, which scales with $\left|x_0^3\right|$).

\section{Comparison with Gaussian Beams\\and Airy Beams Use Cases}\label{Sec: Beam Customization}

\begin{figure*}
  \centering
  \includegraphics[scale=0.34]{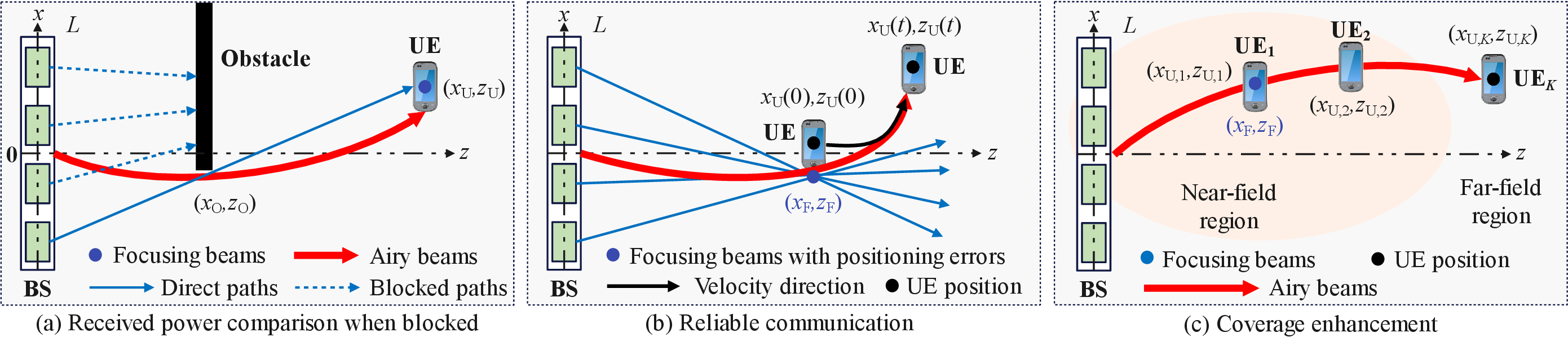}
  \caption{Airy beams compared to focusing beams involving obstacles, and to be robust communication and SE improvement tools.} \label{Fig: beam customization}
\end{figure*}

Airy beams are commonly considered effective in obstruction scenarios \cite{AIScheme, Multiuser}, due to their inherent curved main lobe trajectories and self-healing properties. However, whether they consistently outperform Gaussian beams in such environments still requires further numerical validation.\footnote{In conventional XL-MIMO systems, Gaussian beams refer to focusing beams in the near-field region, where the beamforming vector has a quadratic phase as in {\bl\eqref{Eq: Section3_1_11}} with $x_0 = 0$, thus focusing on a single point. However, Gaussian beams typically refer to steering beams in the far-field region, where the beamforming vector has a linear phase as $2\pi \left(n - \frac{N + 1}{2} \right) d \sin \left(\theta\right)$, for $n = 1,\ldots,N$, thus facilitating a directional beam pointing towards $\theta$ in the far-field region. Note that Airy beams will degenerate into focusing beams when $x_0 = 0$ in the near-field region, while degenerating into steering beams in the far-field region.} In addition, their main lobe trajectories can be flexibly engineered, enabling scenario-dependent use cases in wireless communication. In this section, a quantitative comparison between Airy and Gaussian beams is developed, while scenario-dependent use cases are presented. Note that Airy beams in this section refer to the cubic phase plus focusing phase method \cite{Works3, Works4}.

\subsection{Airy Beams for Circumventing Obstacles}

As shown in {\bl Fig.~\ref{Fig: beam customization}(a)}, such a condition is considered: A user equipment (UE) is located at the position $\left(x_{\rm U}, z_{\rm U}\right)$, while the endpoint of an infinitely long obstacle is located at the position $\left(x_{\rm O}, z_{\rm O}\right)$, where $0 < z_{\rm O} < z_{\rm U}$. Therefore, the path between the $n$-th antenna and the UE is given by
\begin{equation}\label{Eq: Section4_1_1}
x \left( z\right) = \frac{x_{\rm U} - x_n}{z_{\rm U}} z + x_n,
\end{equation}
for $n = 1,\ldots,N$ and $z > 0$, while it will be blocked if $x \left(z_{\rm O}\right) > z_{\rm O}$. Assume that the number of blocked antennas is denoted as $\eta N$ (i.e., from $N - \eta N + 1$ to $N$), where $0 \le \eta \le 1$ is the obstruction ratio. Then, the EF on the position of the UE, $\left( x_{\rm U}, z_{\rm U} \right)$, can be expressed as
\begin{equation}\label{Eq: Section4_1_2}
\psi \left(x_{\rm U}, z_{\rm U} \right) = \psi_{\rm unb} \left(x_{\rm U}, z_{\rm U} \right) + \psi_{\rm b}\left(x_{\rm U}, z_{\rm U} \right),
\end{equation}
where $\psi_{\rm unb} \left(x_{\rm U}, z_{\rm U} \right)$ is the direct EF w.r.t. the unblocked antennas \cite{References9}, such that
\begin{equation}\label{Eq: Section4_1_3}
\psi_{\rm unb} \left(x_{\rm U}, z_{\rm U} \right) = \sum\limits_{n = 1}^{N - \eta N} \psi \left(x_n, 0 \right) \exp \left[ j \frac{k \left(x_{\rm U} - x_n\right)^2}{2z_{\rm U}}\right],
\end{equation}
while $\psi_{\rm b} \left(x_{\rm U}, z_{\rm U} \right)$ is the diffractive EF w.r.t. the blocked antennas, such that
\begin{equation}\label{Eq: Section4_1_4}
\begin{aligned}
&\psi_{\rm b} \left(x_{\rm O}, z_{\rm O}, x_{\rm U}, z_{\rm U} \right) = C_0 \left(x_{\rm O}, z_{\rm O}, x_{\rm U}, z_{\rm U} \right) \\ \times &\sum\limits_{n = N - \eta N+1}^{N} {\psi} \left(x_n,0\right) \exp\left[j \frac{k\left(x_{\rm O} - x_n\right)^2}{2z_{\rm O}} \right],
\end{aligned}
\end{equation}
where $\left|C_0 \left(x_{\rm O}, z_{\rm O}, x_{\rm U}, z_{\rm U} \right)\right| < 1$ is a diffraction coefficient.

\subsubsection{Gaussian Beams}

Assume that Gaussian beams with a focusing beam is adopted in a near-field scenario. Therefore, the beamforming vector for a focusing beam is given by
\begin{equation}\label{Eq: Section4_1_5}
\psi_{\rm G} \left(x_n, 0 \right) = \exp \left[-j \frac{k\left(x_{\rm U} - x_n \right)^2}{2z_{\rm U}} \right].
\end{equation}
Note that when $\eta \to 0$, the diffractive EF can be ignored, since $\left|\psi_{\rm b} \left(x_{\rm U}, z_{\rm U} \right)\right| \ll \left| \psi_{\rm unb} \left(x_{\rm U}, z_{\rm U} \right)\right|$, while the received power of the UE is given by
\begin{equation}\label{Eq: Section4_1_6}
P_{\rm G}|_{\eta \to 0} = \frac{P\lambda^2\left| \psi_{\rm unb} \left(x_{\rm U}, z_{\rm U}\right) \right|^2}{16\pi^2 r_{\rm U}^2N}  = \frac{P\lambda^2\left( 1 - \eta\right)^2 N}{16\pi^2 r_{\rm U}^2},
\end{equation}
where $r_{\rm U} = \sqrt{x_{\rm U}^2 + z_{\rm U}^2}$. However, when $\eta \to 1$, the direct EF can be ignored, since almost the entire array aperture has been blocked, and thus we have
\begin{equation}\label{Eq: Section4_1_7}
P_{\rm G}|_{N \to +\infty}^{\eta \to 1, d \to 0} = \frac{P\lambda^2\left| \psi_{\rm b} \left(x_{\rm O}, z_{\rm O}, x_{\rm U}, z_{\rm U}\right) \right|^2}{16\pi^2 r_{\rm O}^2N} |_{N \to +\infty}^{d \to 0} \to 0,
\end{equation}
which has been demonstrated in near-field beamforming \cite{References10}.

\subsubsection{Airy Beams}

The beamforming vector is set as in {\bl\eqref{Eq: Section3_1_11}} and {\bl\eqref{Eq: Section3_2_1}}, where $\left(x_{\rm F}, z_{\rm F}\right) = \left(x_{\rm O}, z_{\rm O}\right)$, while it is assumed that $\left(x_{\rm U}, z_{\rm U}\right)$ is located on the main lobe trajectory of the Airy beam as in {\bl\eqref{Eq: Section3_1_13}}, {\bl\eqref{Eq: Section3_2_8}}, and {\bl\eqref{Eq: Section3_2_9}}. Note that when $\eta \to 0$, the diffractive EF can be ignored, while the received power of the UE can be expressed as
\begin{equation}\label{Eq: Section4_1_8}
\begin{aligned}
&P_{\rm A} |_{\eta \to 0} = \frac{P \lambda^2}{16\pi^2 r_{\rm U}^2 N}\left|\sum\limits_{n = 1}^{N - \eta N} \exp\left[j\frac{x_0^3}{3} x_n^3 \right]\right|^2 \\ \le &\frac{P \lambda^2}{16\pi^2 r_{\rm U}^2N}\sum\limits_{n = 1}^{N - \eta N} \left|\exp\left[j\frac{x_0^3}{3} x_n^3 \right]\right|^2 = P_{\rm G}|_{\eta \to 0},
\end{aligned}
\end{equation}
where ``='' holds only when $x_0 = 0$ (Airy beams degenerate into focusing beams) or $N = 1$. When $x_0 \ne 0$, we have
\begin{equation}\label{Eq: Section4_1_9}
P_{\rm A} |_{\eta \to 0} =
\left\{ {\begin{array}{*{20}{l}}
{P_{\rm G}|_{\eta \to 0},\:\:\:\:\:\:\:\: \:\:\:\:\:\:\: \:\:\:\, \,\,\,{\rm for} \: N \to 1, \: d \to 0 }\\
{\frac{P\lambda^2 \left(1 - \eta \right)^2}{16 \pi^2 r_{\rm U}^2Nx_0^2d^2} {\rm Ai}_{\rm max}^2, \: {\rm for} \:N \to +\infty, \: d \to 0}
\end{array}} \right.,
\end{equation}
where ${\rm Ai}_{\rm max} = {\rm Ai} \left(\mu_0\right) \approx 0.5356$. However, when $\eta \to 1$, the direct EF can be ignored. Since $\left(x_{\rm F}, z_{\rm F}\right)$ is also located on the main lobe trajectory, we have
\begin{equation}\label{Eq: Section4_1_10}
P_{\rm A} |_{N \to +\infty}^{\eta \to 1, d \to 0} \to \frac{P \lambda^2 \left|C_0 \left(x_{\rm O}, z_{\rm O}, x_{\rm U}, z_{\rm U} \right)\right|^2}{16 \pi^2 r_{\rm O}^2 N x_0^2 d^2} {\rm Ai}_{\rm max}^2.
\end{equation}

\textbf{\textit{Theorem 5:}} When obstruction is minimal, i.e., $\eta$ is small, the perceived obstacle avoidance arises only from main lobe deviation while without compensating for blockage-induced power loss, such that
\begin{equation}\label{Eq: Section4_1_11}
P_{\rm A} |_{\eta \to 0} \ll P_{\rm G} |_{\eta \to 0}.
\end{equation}
However, in scenarios where $\eta \to 1$, the diffraction-dominated paths outperform direct paths. Airy beams achieve higher beamforming gain than focusing beams attributed to the configuration of coherent diffraction phases, i.e.,
\begin{equation}\label{Eq: Section4_1_12}
\frac{P_{\rm A} |_{\eta \to 1}}{P_{\rm G} |_{\eta \to 1}}|_{N \to +\infty}^{d \to 0} \to +\infty.
\end{equation}

\textbf{\textit{Proof:}} See Appendix H.

\textbf{\textit{Proposition 8:}} Note that $\left|\psi_{\rm unb} \left(x_{\rm U}, z_{\rm U}\right)\right|$ of focusing beams is proportional to $N$, while that of Airy beams is not related to $N$. It is evident that the beamforming gain of Airy beams hardly increases as $N$ increases, indicating that the power enhancement advantage of XL-MIMO arrays cannot be optimally leveraged by Airy beams. In a practical XL-MIMO system, the number of antennas used to construct an Airy beam should be set to the minimum value that satisfies {\bl\eqref{Eq: Section3_2_7}}.

\begin{figure}
  \centering
  \includegraphics[scale=0.3]{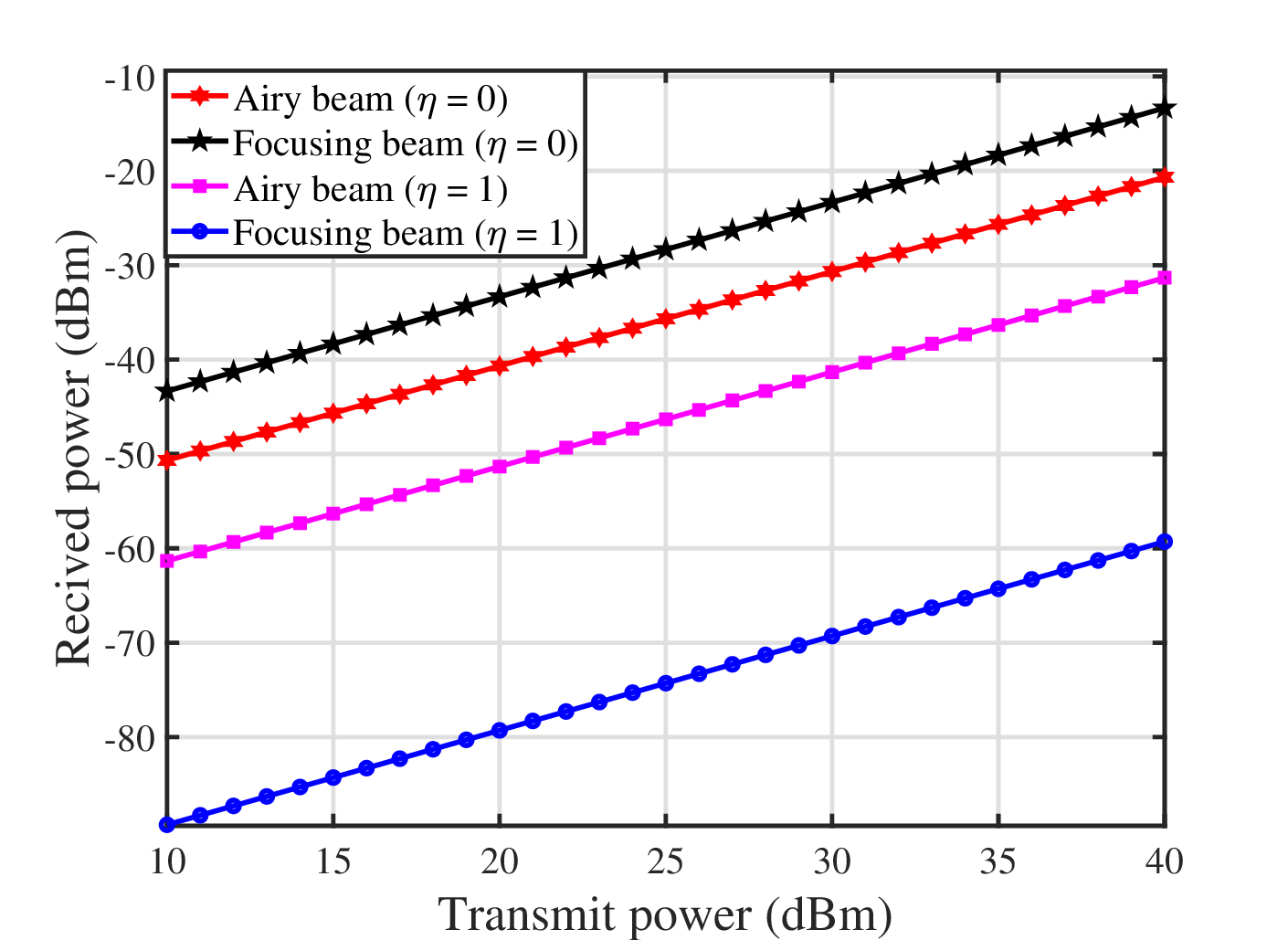}
  \caption{Received power against $P$.} \label{Fig: simulation7}
\end{figure}

The received power against $P$ is evaluated, where $f_{\rm c} = 7$ GHz, $N = 256$, $d = \frac{\lambda}{2}$, $x_0 = 1$, $\left(x_{\rm U}, z_{\rm U} \right) = \left(30, 0 \right)$ m, $x_{\rm O} = 15$ m, and $\left| C_0 \left(x_{\rm O}, z_{\rm O}, x{\rm U}, z_{\rm U} \right) \right|^2 \approx 0.5944$. As shown in {\bl Fig.~\ref{Fig: simulation7}}, focusing beams outperform Airy beams when $\eta \to 0$, while Airy beams become superior when $\eta \to 1$. Moreover, the received power is much compromised with severe obstruction.

\subsection{Scenario-Dependent Use Cases}

Despite the constraints imposed by \textbf{\textit{Proposition 8}}, Airy beams can nevertheless be employed to enhance near-field communication performance in certain scenarios in comparison with Gaussian beams.

\subsubsection{Robust Communication}

Gaussian beams with a focusing beam have been demonstrated to achieve significant power enhancement \cite{BeamFocusing}, while the beamforming gain is also highly sensitive to positioning errors and UE's mobility, which will compromise the near-field communication robustness \cite{References11}. Compared to focusing beams, Airy beams exhibit a comparatively wider beamwidth and a non-diffractional main lobe trajectory, thus providing the potential of enhancing robustness. Assume that the initial position and velocity of a UE are $\left(x_{\rm U} \left(0\right), z_{\rm U} \left(0\right) \right)$ and $\left(v_{{\rm U}, x}, v_{{\rm U}, z}\right)$, where $z_{\rm U} \left(0\right) > 0$, while the real-time position of the UE is denoted as
\begin{equation}\label{Eq: Section4_2_1}
\left(x_{\rm U} \left(t\right), z_{\rm U} \left(t\right)\right) = \left(x_{\rm U} \left(0\right), z_{\rm U} \left(0\right)\right) + \left(v_{x}, v_{z}\right) t,
\end{equation}
for $0 \le t \le {\cal T}$, where ${\cal T} > 0$ is the beam update time interval.

$\bullet$ \textbf{\textit{SE against positioning errors:}} In the initial time when $t = 0$, assume that the positioning errors along the $x$-axis and $z$-axis are indicated by $\delta x$ and $\delta z$, respectively. The positioning results is denoted as $\left(\hat x_{\rm U} \left(0 \right), \hat z_{\rm U} \left(0 \right)\right) = \left(x_{\rm U} \left(0\right), z_{\rm U} \left(0\right)\right) + \left(\delta x, \delta z \right)$, and thus the beamforming vector is set as in {\bl\eqref{Eq: Section3_1_11}} and {\bl\eqref{Eq: Section3_2_1}}, where $\left(x_{\rm F}, z_{\rm F}\right) = \left(\hat x_{\rm U} \left(0\right), \hat z_{\rm U}\left(0\right) \right)$. Then, the received power of the UE can be expressed as
\begin{equation}\label{Eq: Section4_2_2}
{\bar P}_{\rm A} = {\frac{P\lambda^2}{16\pi^2r_{\rm U}^2N} \left|\sum\limits_{n = 1}^N {\exp\left[j \phi \left(x_n, 0 \right) \right]}\right|^2},
\end{equation}
where
\begin{equation}\label{Eq: Section4_2_3}
\begin{aligned}
&\phi \left(x_n, 0 \right) \\= &\frac{x_0^3}{3}x_n^3 - \frac{k\left(x_{\rm U}\left(0 \right) + \delta x - x_n\right)^2}{2z_{\rm U}\left(0 \right) + \delta z} + \frac{k \left(x_{\rm U}\left(0 \right) - x_n\right)^2}{2z_{\rm U}\left(0 \right)},
\end{aligned}
\end{equation}
for $n = 1, \ldots, N$, while
\begin{equation}\label{Eq: Section4_2_4}
\left| x_0\right| > \frac{k}{z_{\rm U} \left(0\right) \left| \mu_1 - \mu_0\right|} \left|\delta x\right|.
\end{equation}
Note that the beamwidth of an Airy beam is $\left| \mu_1 - \mu_0\right|$, where $\mu_1 \approx -2.338$ is the first zero point of ${\rm Ai} \left(x\right)$.

$\bullet$ \textbf{\textit{SE against UE’s mobility:}} Assume that $\delta x = 0$ and $\delta z = 0$. As shown in {\bl Fig.~\ref{Fig: beam customization}(b)}, the main lobe trajectory of an Airy beam in XL-MIMO systems can be designed according to {\bl\eqref{Eq: Section3_1_13}} to be along the UE’s mobility direction, thus enhancing the received power of the UE. The beamforming vector is set as in {\bl\eqref{Eq: Section3_1_11}} and {\bl\eqref{Eq: Section3_2_1}}, where $\left(x_{\rm F}, z_{\rm F}\right) = \left(x_{\rm U} \left(0\right), z_{\rm U}\left(0\right) \right)$. Then, the received power of the UE is given by
\begin{equation}\label{Eq: Section4_2_5}
{\bar P}_{\rm A} = \frac{1}{\cal T} \int_{t = 0}^{\cal T} {\frac{P\lambda^2}{16\pi^2r_{\rm U}^2N} \left|\sum\limits_{n = 1}^N {\exp\left[j \phi \left(x_n, t \right) \right]}\right|^2} dt,
\end{equation}
where
\begin{equation}\label{Eq: Section4_2_6}
\phi \left(x_n, t\right) = \frac{x_0^3}{3}x_n^3 - \frac{k\left(x_{\rm U}\left(0 \right) - x_n\right)^2}{2z_{\rm U}\left(0 \right)} + \frac{k \left(x_{\rm U}\left(t\right) - x_n\right)^2}{2z_{\rm U}\left(t\right)},
\end{equation}
where $z_{\rm U}\left(t\right) > 0$, for $n = 1,\ldots,N$, and $0 \le t \le {\cal T}$. Specifically, $x_0$ should be set to satisfy
\begin{equation}\label{Eq: Section4_2_7}
x_0 = \arg\min_{x_0} \left|\frac{\partial x \left(z\right)}{\partial z} - \frac{v_{{\rm U}, x}}{v_{{\rm U}, z}}\right|.
\end{equation}

These representative examples demonstrate the potential of Airy beams to enhance near-field communication robustness in 6G. In particular, they are well suited for low-altitude and highly dynamic environments, where inevitable positioning errors and large velocity pose significant challenges to conventional beamforming techniques.

\begin{figure}
  \centering
  \includegraphics[scale=0.3]{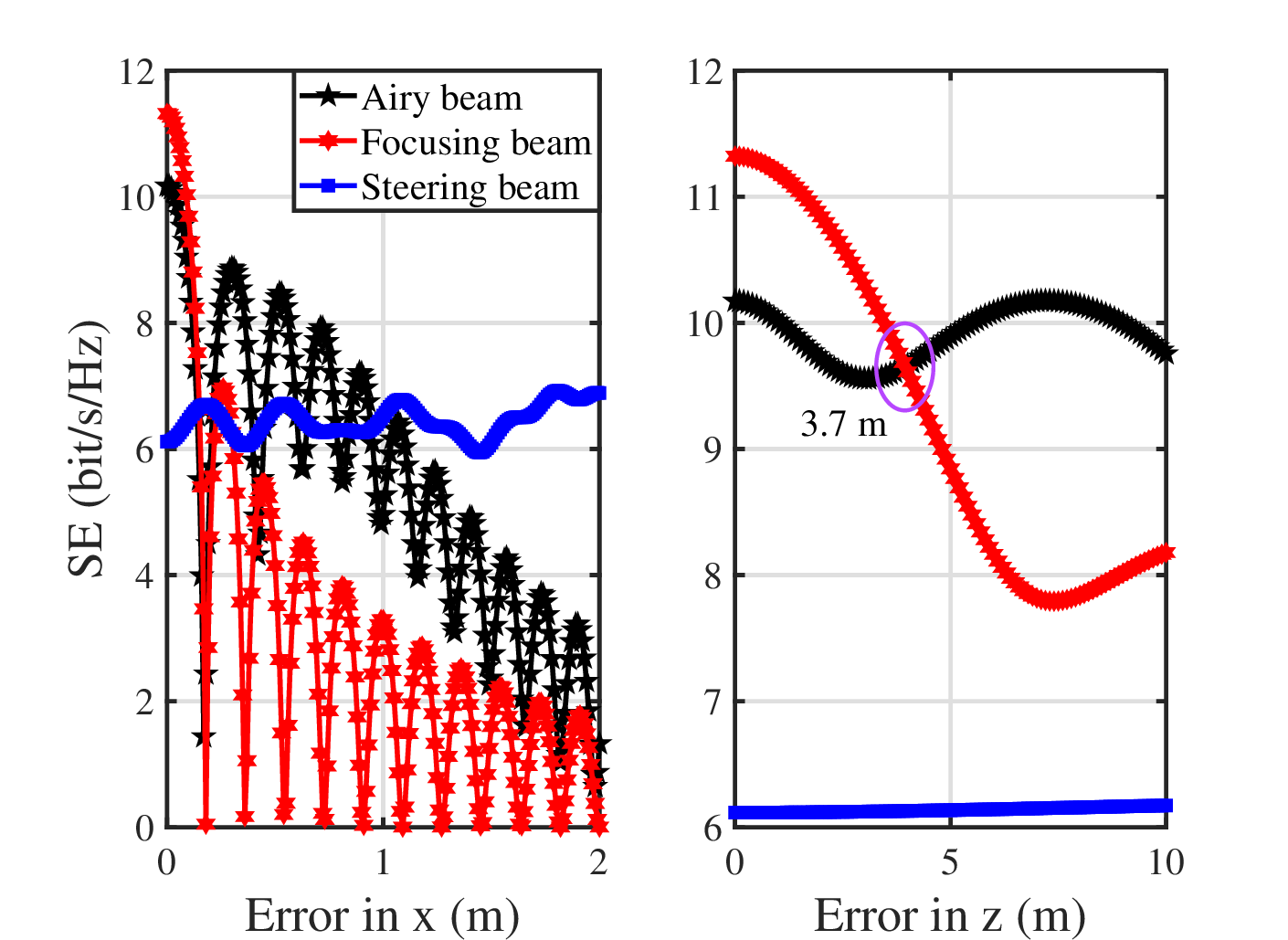}
  \caption{SE against positioning errors $\left|\delta x \right|$ and $\left|\delta z \right|$, respectively.} \label{Fig: simulation4}
\end{figure}

\begin{figure}
  \centering
  \includegraphics[scale=0.3]{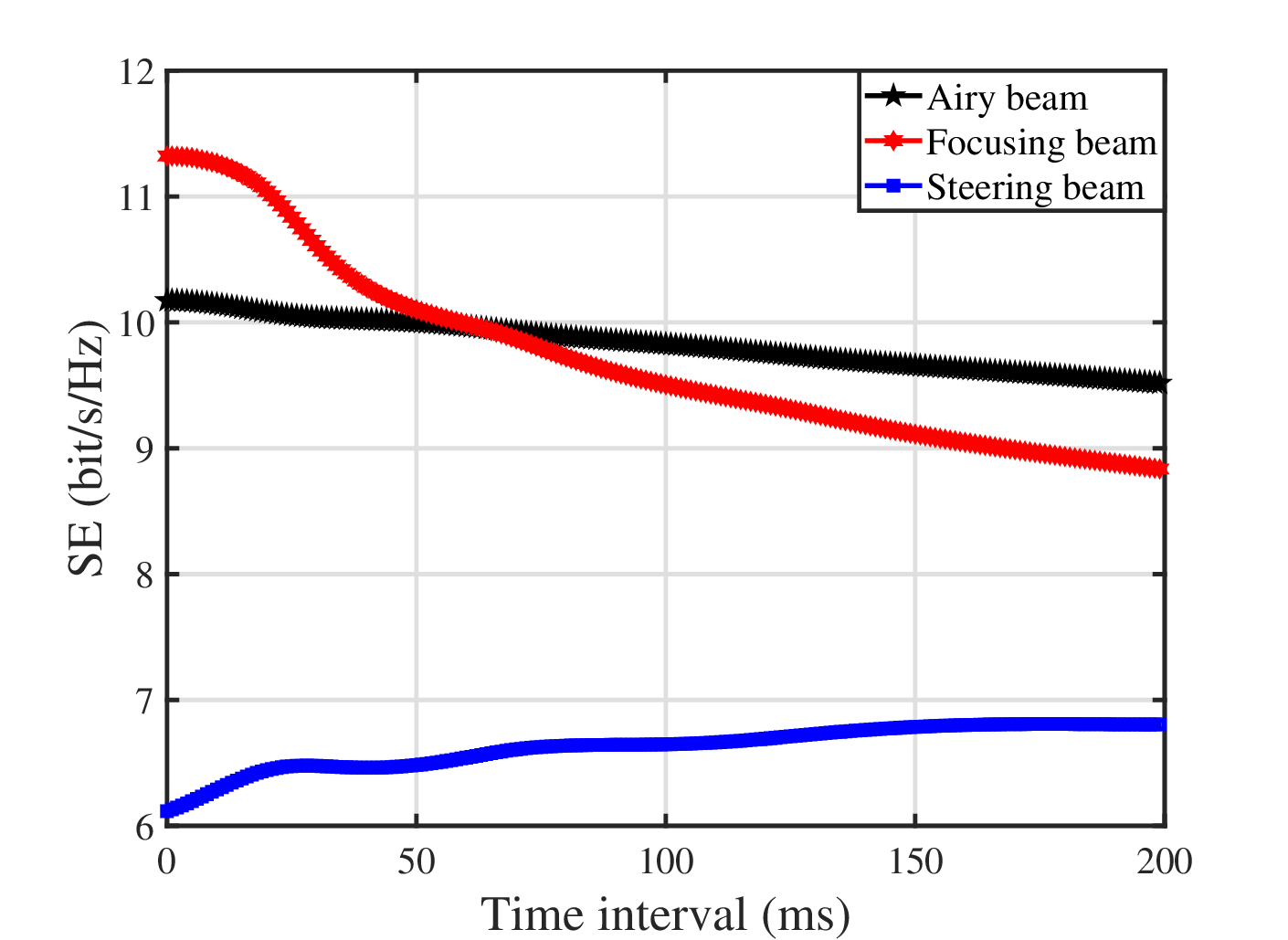}
  \caption{SE against time interval $\cal T$.} \label{Fig: simulation5}
  \vspace{-0.4cm}
\end{figure}

The SE is evaluated against $\left|\delta x \right|$, $\left|\delta z \right|$, and $\cal T$, where $f_{\rm c} = 7$ GHz, $N = 256$, $d = \frac{\lambda}{2}$, $x_0 = 1$, $\left(x_{\rm F}, z_{\rm F}\right) = \left(0, 20 \right)$ m, and the signal-to-noise (SNR) is 10 dB. The SE is defined as
\begin{equation}\label{Eq: Section4_2_8}
{\rm SE} = \log_{10} \left( 1 + \frac{\bar P_{\rm A}}{\sigma^2} \right),
\end{equation}
where $\sigma^2$ is the variance of a complex additive Gaussian white noise (which has zero mean). As shown in {\bl Fig.~\ref{Fig: simulation4}}, when the UE position is perfectly known, Gaussian beams with beam focusing phases achieve the highest SE. However, the SE of focusing beams \cite{BeamFocusing} degrades rapidly in the presence of inevitable positioning errors. In contrast, Airy beams exhibit enhanced robustness under such imperfections. Note that both Airy beams and focusing beams are sensitive to $\left|\delta x\right|$. When $\left|\delta x\right|$ becomes large, steering beams \cite{SteeringBeam} achieve superior SE. As shown in {\bl Fig.~\ref{Fig: simulation5}}, Airy beams also exhibit improved robustness in highly dynamic scenarios with a velocity 100 m/s. As $\mathcal{T}$ increases, the SE of Airy beams degrades more slowly compared to that of focusing beams.

\subsubsection{SE Improvement}

As shown in {\bl Fig.~\ref{Fig: beam customization}(c)}, the non-diffraction property of Airy beams provides the potential to serve multiple UEs SE by a single beam. Compared to focusing beams that focus on only a single point, and steering beams suffering from issues of power diffusion in near-field regions \cite{References12}, Airy beams achieve a trade-off between beamforming gain and the served UE number in certain scenarios.

Assume that the BS simultaneously serves $K$ UEs by an Airy beam using frequency division multiplexing (in XL-MIMO systems with hybrid precoding architectures), while the position of each UE, $\left(x_{{\rm U}, k}, z_{{\rm U},k} \right)$, is located on the main lobe trajectory of Airy beams, where $z_{{\rm U},k} > 0$, for $k = 1,\ldots,K$. To achieve equilibrium in each UE's quality of service, the transmit power for the $k$-th UE is set as
\begin{equation}\label{Eq: Section4_2_9}
P_{{\rm U}, k} = \frac{P r_{{\rm U}, k}^2}{\sum\nolimits_{k = 1}^K r_{{\rm U}, k}^2},
\end{equation}
where $r_{{\rm U},k} = \sqrt{x_{{\rm U},k} + z_{{\rm U},k}^2}$, such that $\frac{P_{{\rm U},1}}{r_{{\rm U},1}^2} = \frac{P_{{\rm U},2}}{r_{{\rm U},2}^2} = \ldots = \frac{P_{{\rm U},K}}{r_{{\rm U},K}^2}$, and $\sum\nolimits_{k = 1}^K P_{{\rm U},k} = P$. Subsequently, the average SE of the XL-MIMO system with $K$ UEs is defined as
\begin{equation}\label{Eq: Section4_2_10}
\begin{aligned}
& {\rm SE} \\ = & \frac{1}{K} \sum\limits_{k = 1}^{K} \log_2 \left(1 + \frac{P_{{\rm U}, k}\lambda^2}{16 \pi^2 r_{{\rm U}, k}^2 N \sigma^2} \left| \sum\limits_{n = 1}^N \exp\left[j \phi \left(x_n, k \right)\right] \right|^2 \right),
\end{aligned}
\end{equation}
where
\begin{equation}\label{Eq: Section4_2_11}
\phi\left(x_n, k\right) = \frac{x_0^3}{3} x_n^3 - \frac{k \left(x_{{\rm U}, 1} - x_n \right)^2}{2z_{{\rm U}, 1}} + \frac{k\left(x_{{\rm U}, k} - x_n \right)^2}{2z_{{\rm U}, k}},
\end{equation}
for $n = 1,\ldots,N$, where the focusing point is $\left(x_{{\rm U}, 1}, z_{{\rm U}, 1} \right)$.

\begin{figure}
  \centering
  \includegraphics[scale=0.3]{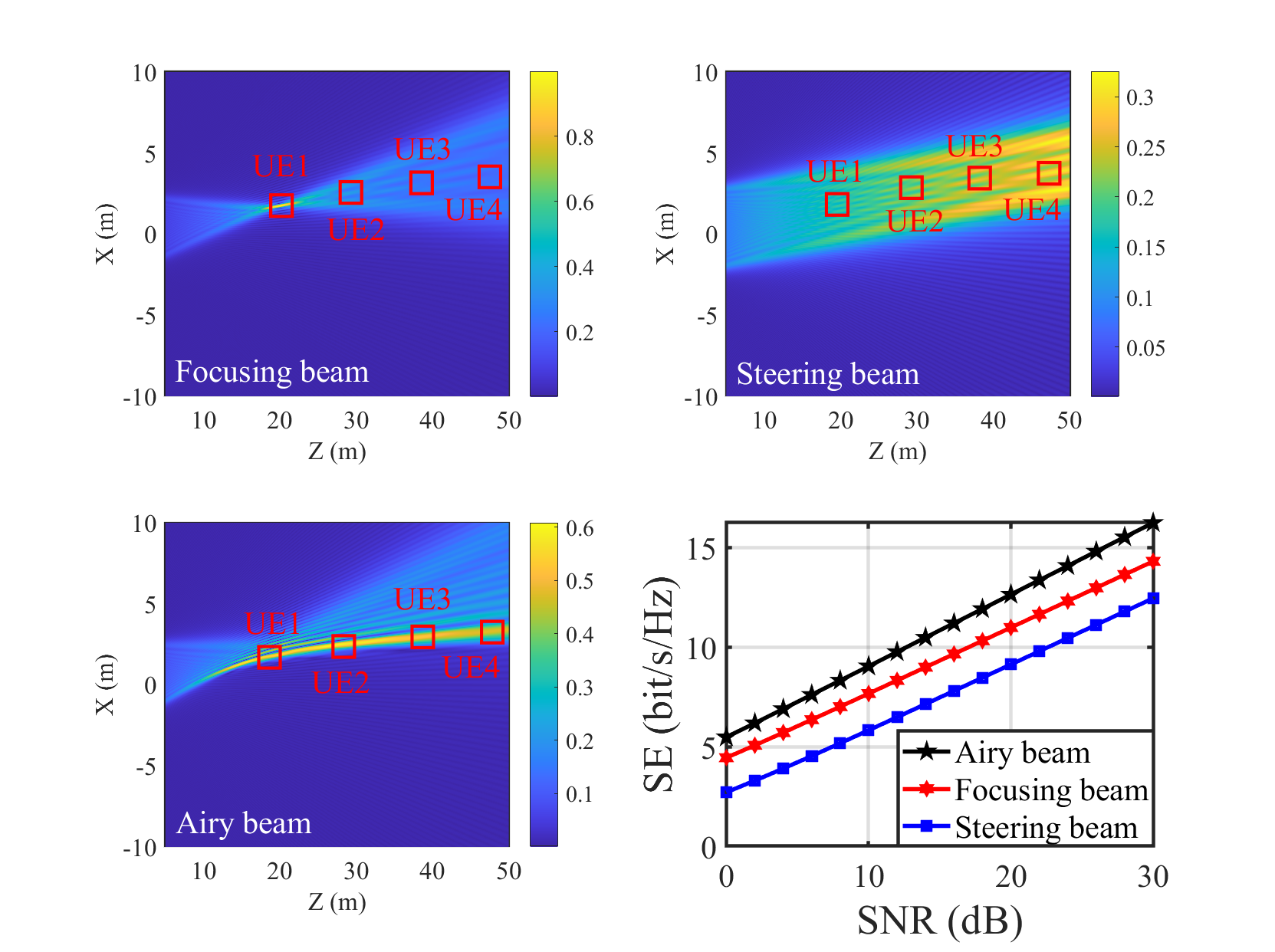}
  \caption{SE against the SNR with Airy, focusing, and steering beams.} \label{Fig: simulation8}
  \vspace{-0.4cm}
\end{figure}

As shown in {\bl Fig.~\ref{Fig: simulation8}}, the SE of the XL-MIMO system against the SNR is evaluated, where $f_{\rm c} = 7$ GHz, $N = 256$, $d = \frac{\lambda}{2}$, $x_0 = 1.1$, and $K = 4$. In certain scenarios where the main lobe trajectory of an Airy beam is designed to be along the positions of multiple UEs, it achieves a higher SE than that of focusing beams and steering beams, due to the non-diffractional propagation property in the near-field region.

\section{Conclusion}\label{Sec: Conclusion}

This paper established an analytical theoretical framework of Airy beams in XL-MIMO systems. The actual EFs and main lobe trajectories were derived considering the cubic phase method and the cubic phase plus focusing phase method, while the constraints on array aperture and antenna spacing for the ``truncation'' and ``sampling'' operations were determined. In addition, quantitative metrics were developed to rigorously evaluate the performance trade-offs between Airy beams and near-field focusing beams, indicating that Airy beams achieve superior performance in scenarios with severe obstruction, while suboptimal in scenarios where obstruction is minimal. The scenario-dependent use cases of Airy beams were also presented to enhance robustness and improve SE in wireless communication. Numerical results corroborated the proposed analytical theoretical framework of Airy beams in near-field XL-MIMO, and proved Airy beams to be robustness and SE enhancement tools.

\begin{appendices}

\section{Proof of Theorem 1}

Substituting {\bl\eqref{Eq: Section3_1_3}} into {\bl\eqref{Eq: Section3_1_2}}, the EF on the position $\left(x,z\right)$, for any $z > 0$, can be calculated according to
\begin{equation}\label{Eq: SectionA_1}
\begin{aligned}
&{\psi}_{\rm I} \left(x,z\right) = \int_{-\infty}^{+\infty} {\exp \left[j\left(\frac{x_0^3}{3}x'^3 + \frac{k \left(x - x'\right)^2}{2z}\right]\right)} dx' \\
= &\int_{-\infty}^{+\infty} {\exp \left[j\left( \frac{x_0^3}{3} x'^3 + \frac{k}{2z}x'^2 - \frac{kx}{z}x' + \frac{kx^2}{2z} \right) \right]} dx' \\
\mathop = \limits^{\left(\rm b \right)} & \frac{1}{x_0} \int_{-\infty}^{+\infty} {\exp \left[j\left( \frac{1}{3}t^3 + \frac{k}{2zx_0^2}t^2 - \frac{kx}{zx_0}t + \frac{kx^2}{2z} \right) \right]} dt,
\end{aligned}
\end{equation}
where (b) is derived by $t = x_0 x'$. Subsequently, the formula {\bl\eqref{Eq: SectionA_1}} can be alternatively expressed as
\begin{equation}\label{Eq: SectionA_2}
\begin{aligned}
&{\psi}_{\rm I} \left(x,z\right) = \frac{1}{x_0} \exp \left[j \phi_1 \right] \\ \times &\int_{-\infty}^{+\infty} {\exp \left[j\left( \frac{1}{3}\left(t + \frac{k}{2zx_0^2} \right)^3 - \frac{k^2}{4z^2x_0^4}t - \frac{kx}{zx_0}t \right) \right]} dt \\
\mathop = \limits^{\left(\rm c \right)} & \frac{1}{x_0} \int_{-\infty}^{+\infty} {\exp \left[j\left( \frac{1}{3}t'^3 + \left(-\frac{k^2}{4z^2x_0^4} - \frac{kx}{zx_0} \right)t' + \phi_2 \right) \right]} dt' \\
= & \frac{1}{x_0} {\rm Ai} \left( -\frac{k^2}{4z^2 x_0^4} - \frac{kx}{zx_0} \right) \exp \left[j \phi_2 \right],
\end{aligned}
\end{equation}
where (c) is derived by $t' = t + \frac{k}{2 z x_0^2}$, while
\begin{equation}\label{Eq: SectionA_3}
\phi_1 = -\frac{k^3}{24z^3x_0^6} + \frac{kx^2}{2z},
\end{equation}
and
\begin{equation}\label{Eq: SectionA_4}
\phi_2 = \frac{k^3}{12z^3x_0^6} + \frac{k^2 x}{2 z^2 x_0^3} + \frac{kx^2}{2z}.
\end{equation}
Consequently, the main lobe trajectory of the Airy beam in {\bl\eqref{Eq: SectionA_2}} can be calculated according to
\begin{equation}\label{Eq: SectionA_5}
 -\frac{k^2}{4z^2 x_0^4} - \frac{kx}{zx_0} = \mu_0,
\end{equation}
such that
\begin{equation}\label{Eq: SectionA_6}
x\left(z\right) = -\frac{k}{4 x_0^3} \frac{1}{z} - \frac{\mu_0x_0}{k}z.
\end{equation}
Therefore, \textbf{\textit{Theorem 1}} is proved. \qed

\section{Proof of Theorem 2}

\begin{figure*}[ht]
\centering
\begin{equation}\label{Eq: SectionB_2}
\begin{aligned}
{\psi} \left(x,z\right) = &\int_{-\infty}^{+\infty} {\exp \left[j \left( \frac{x_0^3}{3} x'^3 + \frac{k}{2z}x'^2 - \frac{k}{2z_{\rm F}}x'^2 - \frac{kx}{z}x' + \frac{kx_{\rm F}}{z_{\rm F}}x' + \frac{kx^2}{2z} - \frac{kx_{\rm F}^2}{2z_{\rm F}} \right) \right]} dx' \\
\mathop = \limits^{\left(\rm b\right)} & \frac{1}{x_0} \exp \left[j \left(\frac{kx^2}{2z} - \frac{kx_{\rm F}^2}{2z_{\rm F}} \right) \right] \int_{-\infty}^{+\infty} {\exp \left[j\left( \frac{1}{3}t^3 + \frac{k}{2x_0^2}\left(\frac{1}{z} - \frac{1}{z_{\rm F}} \right)t^2 - \frac{k}{x_0} \left(\frac{x}{z} - \frac{x_{\rm F}}{z_{\rm F}} \right) t \right) \right]} dt \\
= & \frac{1}{x_0} \exp \left[j \phi_3 \right] \int_{-\infty}^{+\infty} {\exp \left[j\left( \frac{1}{3}\left(t + \frac{k\left(z_{\rm F} - z\right)}{2zz_{\rm F}x_0^2} \right)^3 - \frac{k^2\left(z_{\rm F} - z\right)^2}{4z^2z_{\rm F}^2 x_0^4} t - \frac{k \left(xz_{\rm F} - x_{\rm F}z\right)}{z z_{\rm F} x_0} t \right) \right]} dt \\
\mathop = \limits^{\left(\rm d \right)} & \frac{1}{x_0} \exp \left[j \phi_4 \right] \int_{-\infty}^{+\infty} {\exp \left[j\left( \frac{1}{3}t'^3 + \left( - \frac{k^2 \left(z_{\rm F} - z \right)^2}{4z^2 z_{\rm F}^2 x_0^4} - \frac{k\left(xz_{\rm F} - x_{\rm F}z\right)}{z z_{\rm F} x_0} \right)t'\right) \right]} dt' \\
= & \frac{1}{x_0} {\rm Ai} \left( - \frac{k^2 \left(z_{\rm F} - z \right)^2}{4 z^2 z_{\rm F}^2 x_0^4} -\frac{k\left(xz_{\rm F} - x_{\rm F}z \right)}{zz_{\rm F} x_0} \right) \exp \left[j \phi_4 \right]
\end{aligned}
\end{equation}
\hrule
\end{figure*}

Substituting {\bl\eqref{Eq: Section3_1_11}} into {\bl\eqref{Eq: Section3_1_10}}, the EF on the position $\left(x, z\right)$, for any $z > 0$, can be calculated by
\begin{equation}\label{Eq: SectionB_1}
\begin{aligned}
&{\psi}_{\rm I} \left(x,z\right) = \\ &\int_{-\infty}^{+\infty} {\exp \left[j \left(\frac{x_0^3}{3} x'^3 - \frac{k \left(x_{\rm F} - x' \right)^2}{2z_{\rm F}} + \frac{k \left(x - x'\right)^2}{2z}\right) \right]} dx',
\end{aligned}
\end{equation}
which can be further calculated as {\bl\eqref{Eq: SectionB_2}}. Note that (b) is derived by $t = x_0 x'$, while (d) is derived by
\begin{equation}\label{Eq: SectionB_3}
t' = t + \frac{k \left(z_{\rm F} - z\right)}{2 z z_{\rm F}x_0^2}.
\end{equation}
Moreover, $\phi_3$ can be expressed as
\begin{equation}\label{Eq: SectionB_4}
\phi_3 = -\frac{k^3 \left(z_{\rm F} - z\right)^3}{24 z^3 z_{\rm F}^3 x_0^6} + \frac{kx^2}{2z} - \frac{kx_{\rm F}^2}{2z_{\rm F}},
\end{equation}
while $\phi_4$ can be denoted as
\begin{equation}\label{Eq: SectionB_5}
\phi_4 = \frac{k^3 \left(z_{\rm F} - z\right)^3}{12z^3 z_{\rm F}^3 x_0^6} + \frac{k^2 \left(xz_{\rm F} - x_{\rm F}z \right) \left(z_{\rm F} - z \right)}{2z^2 z_{\rm F}^2 x_0^3} + \frac{kx^2}{2z} - \frac{kx_{\rm F}^2}{2 z_{\rm F}}.
\end{equation}
Consequently, the main lobe trajectory of the Airy beam in {\bl\eqref{Eq: SectionB_2}} can be calculated according to
\begin{equation}\label{Eq: SectionB_6}
- \frac{k^2 \left(z_{\rm F} - z \right)^2}{4 z^2 z_{\rm F}^2 x_0^4} - \frac{k\left(xz_{\rm F} - x_{\rm F}z \right)}{zz_{\rm F} x_0} = \mu_0,
\end{equation}
such that
\begin{equation}\label{Eq: SectionB_7}
x\left(z\right) = - \frac{k}{4x_0^3} \frac{1}{z} + \left( \frac{x_{\rm F}}{z_{\rm F}} - \frac{k}{4z_{\rm F}^2x_0^3} - \frac{\mu_0x_0}{k} \right) z + \frac{k}{2z_{\rm F}x_0^3}.
\end{equation}
Hence, \textbf{\textit{Theorem 2}} has been proved. \qed

\section{Proof of Proposition 3}

For the sake of illustration, only $x_0 > 0$ is considered, while a similar conclusion can be obtained for $x_0 < 0$. Note that $\frac{\partial^2 x \left(z\right)}{\partial x^2} < 0$ according to {\bl\eqref{Eq: Section3_1_15}}, while Part 4 always exceeds 0 with monotonically increasing from $0$ to $+\infty$ along $z$, for $z > 0$. Therefore, when Part 5 is less than 0, i.e.,
\begin{equation}\label{Eq: SectionC_1}
\frac{x_{\rm F}}{z_{\rm F}} - \frac{k}{4 z_{\rm F}^2 x_0^3} - \frac{\mu_0 x_0}{k} < 0,
\end{equation}
the formula {\bl\eqref{Eq: Section3_1_14}} has a root at $z_{\rm ext}$, such that
\begin{equation}\label{Eq: SectionC_2}
\frac{k}{4x_0^3} \frac{1}{z_{\rm ext}^2} + \frac{x_{\rm F}}{z_{\rm F}} - \frac{k}{4 z_{\rm F}^2 x_0^3} - \frac{\mu_0 x_0}{k} = 0.
\end{equation}
Hence, based on {\bl\eqref{Eq: SectionC_1}} and {\bl\eqref{Eq: SectionC_2}}, \textbf{\textit{Proposition 3}} is proved. \qed

\section{Proof of Theorem 3}

In particular, the formula {\bl\eqref{Eq: Section3_1_10}} is an oscillation integral. When the beamforming vector is set as {\bl\eqref{Eq: Section3_1_11}}, the phase function in {\bl\eqref{Eq: Section3_1_10}} can be defined as
\begin{equation}\label{Eq: SectionD_1}
\phi \left(x' \right) = \frac{x_0^3}{3} x'^3 - \frac{k \left(x_{\rm F} - x'\right)^2}{2z_{\rm F}} + \frac{k\left(x - x'\right)^2}{2z},
\end{equation}
where $z > 0$. Then, according to \textbf{\textit{Lemma 1}}, the second-order stationary point (i.e., the caustics) can be calculated by
\begin{equation}\label{Eq: SectionD_2}
\phi' \left(x' \right) = x_0^3 x'^2 + \frac{k\left(z_{\rm F} - z\right)}{zz_{\rm F}} x' - \frac{k\left(xz_{\rm F} - x_{\rm F}z\right)}{z z_{\rm F}} = 0,
\end{equation}
and
\begin{equation}\label{Eq: SectionD_3}
\phi'' \left(x' \right) = 2x_0^3 x' + \frac{k\left(z_{\rm F} - z\right)}{zz_{\rm F}} = 0.
\end{equation}
Then, the second-order stationary point is calculated by {\bl\eqref{Eq: SectionD_3}} and expressed as
\begin{equation}\label{Eq: SectionD_4}
x_{\rm c}' = - \frac{k }{2 x_0^3} \frac{1}{z} + \frac{k}{2 z_{\rm F} x_0^3}.
\end{equation}
Substituting {\bl\eqref{Eq: SectionD_4}} into {\bl\eqref{Eq: SectionD_2}}, the caustics is given by
\begin{equation}\label{Eq: SectionD_5}
x\left(z\right) = -\frac{k}{4 x_0^3} \frac{1}{z} + \left( \frac{x_{\rm F}}{z_{\rm F}} - \frac{k}{4z_{\rm F}^2 x_0^3} \right) z + \frac{k}{2 z_{\rm F} x_0^3}.
\end{equation}
Note that by considering the offset $\mu_0$ of the Airy function (i.e., the offset between the caustics and the main lobe), the main lobe trajectory as in {\bl\eqref{Eq: Section3_1_13}} can be obtained. Since $x_{\rm c}'$ must be within the array aperture, we have
\begin{equation}\label{Eq: SectionD_6}
\max_{z} \left|x_{\rm c}'\right| = \frac{k}{2 \left|x_0^3\right|} \left|\frac{1}{z_{\rm F}} - \frac{1}{z} \right| \le \frac{L}{2}.
\end{equation}
Furthermore, it is evident that the main lobe trajectory is contributed not only by the second-order stationary point, but also by the area surrounding $x_{\rm c}'$. Note that
\begin{equation}\label{Eq: SectionD_7}
\phi \left( x' \right) = \phi \left( x_{\rm c}' \right) + \frac{\xi^3}{6} \phi''' \left( x_{\rm c}' \right) = \phi \left( x_{\rm c}' \right) + \frac{x_0^3}{3} \xi^3,
\end{equation}
since $\phi'' \left( x_{\rm c}' \right) = 0$ and $\phi'' \left( x_{\rm c}' \right) = 0$, where $\xi = x' - x_{\rm c}'$. Moreover, the third-order partial derivative of $\phi \left(x'\right)$ with respect to $x'$ is $\phi''' \left(x' \right) = 2x_0^3$, while the fourth-order partial derivative of $\phi \left(x'\right)$ with respect to $x'$ is $\phi'''' \left(x' \right) = 0$. The effective range typically satisfies $\left| \frac{x_0^3}{3} \xi^3\right| \le 1$, and thus
\begin{equation}\label{Eq: SectionD_8}
\left| \xi \right| \le \left|3 x_0^{-3}\right|^{\frac{1}{3}}.
\end{equation}
Then, the array aperture should satisfy
\begin{equation}\label{Eq: SectionD_9}
\max_{z} \left|x_{\rm c}'\right| + \left|3 x_0^{-3}\right|^{\frac{1}{3}} \le \frac{L}{2}.
\end{equation}
Substituting {\bl\eqref{Eq: SectionD_4}} into {\bl\eqref{Eq: SectionD_9}}, and then \textbf{\textit{Theorem 3}} is proved. \qed

\section{Proof of Proposition 5}

For the sake of illustration, only $x_0 > 0$ is considered, while a similar conclusion can be obtained for $x_0 < 0$. The truncation error in {\bl\eqref{Eq: Section3_2_11}} can be alternatively expressed as
\begin{equation}\label{Eq: SectionE_1}
\begin{aligned}
&\left|\epsilon_1 \left(x, z\right) \right| \\ \le &\left|\int_{\frac{L}{2}}^{+\infty} {\exp \left[j\phi \left(x' \right)\right]} dx'\right| + \left|\int_{-\infty}^{-\frac{L}{2}} {\exp \left[j\phi \left(x' \right)\right]} dx' \right|,
\end{aligned}
\end{equation}
for $z > 0$. The phase function $\phi \left(x'\right)$ and $\phi' \left(x'\right)$ has been defined as {\bl\eqref{Eq: SectionD_1}} and {\bl\eqref{Eq: SectionD_2}}, while {\bl\eqref{Eq: SectionD_2}} is a quadratic function of $x'$. When $\left(x, z\right)$ is located on the main lobe trajectory and for $z_{\rm min} \le z \le z_{\rm max}$, the root of {\bl\eqref{Eq: SectionD_1}} and {\bl\eqref{Eq: SectionD_2}} (i.e., the second-order stationary point) is located within the array aperture as in {\bl\eqref{Eq: SectionD_9}} according to \textbf{\textit{Theorem 3}}. Hence, on the main lobe trajectory, $\phi' \left(x'\right)$ monotonically increases for $x' \ge \frac{L}{2}$, while monotonically decreasing for $x' \le \frac{L}{2}$. Moreover, $\phi' \left(x'\right) > 0$ for $\left|x' \right| \ge \frac{L}{2}$. Then, we have
\begin{equation}\label{Eq: SectionE_2}
\begin{aligned}
&\min_{x' \ge \frac{L}{2}} \left| \phi' \left(x' \right) \right| = \left| \phi' \left(\frac{L}{2} \right) \right| \\ = &\frac{\left|zz_{\rm F}x_0^3L^2 + 2k\left(z_{\rm F} - z\right)L - 4k\left(xz_{\rm F} - x_{\rm F}z\right) \right|}{4zz_{\rm F}} \sim {\cal O} \left(L^2\right),
\end{aligned}
\end{equation}
and
\begin{equation}\label{Eq: SectionE_3}
\begin{aligned}
&\min_{x' \le -\frac{L}{2}} \left| \phi' \left(x' \right) \right| = \left| \phi' \left(- \frac{L}{2} \right) \right| \\ = &\frac{\left|zz_{\rm F}x_0^3L^2 - 2k\left(z_{\rm F} - z\right)L - 4k\left(xz_{\rm F} - x_{\rm F}z\right) \right|}{4zz_{\rm F}} \sim {\cal O} \left(L^2\right).
\end{aligned}
\end{equation}
According to \textbf{\textit{Lemma 2}}, the formula {\bl\eqref{Eq: SectionE_1}} can be denoted as
\begin{equation}\label{Eq: SectionE_4}
\begin{aligned}
\left|\epsilon_1 \left(x, z\right) \right| \le &\frac{2}{\left| \phi' \left(\frac{L}{2} \right) \right|} + \frac{2}{\left| \phi' \left(- \frac{L}{2} \right) \right|} \\ \sim & {\cal O} \left(L^{-2}\right) + {\cal O} \left(L^{-2}\right) \sim {\cal O} \left(L^{-2}\right).
\end{aligned}
\end{equation}
Therefore, \textbf{\textit{Proposition 5}} is proved. \qed

\section{Proof of Theorem 4}

According to Taylor series expansion, we have
\begin{equation}\label{Eq: SectionF_1}
\phi \left(x' + d \right) = \phi \left(x' \right) + d \phi' \left(x'\right) + \frac{d^2}{2} \phi'' \left(x'\right) + \frac{d^3}{6} \phi''' \left(x'\right),
\end{equation}
where $\phi'''' \left(x'\right) = 0$, and
\begin{equation}\label{Eq: SectionF_2}
\begin{aligned}
\phi' \left(x' + d \right) = &\phi' \left(x' \right) + d \phi'' \left(x'\right) + \frac{d^2}{2} \phi''' \left(x'\right).
\end{aligned}
\end{equation}
Substituting {\bl\eqref{Eq: SectionF_1}} into {\bl\eqref{Eq: Section3_2_13}}, $D_{\rm d}^{\left(1\right)} \phi \left(x'\right)$ can be denoted as
\begin{equation}\label{Eq: SectionF_3}
D_{\rm d}^{\left(1\right)} \phi \left(x'\right) = \phi' \left(x'\right) + \frac{d^2}{6} \phi''' \left(x'\right).
\end{equation}
Then, substituting {\bl\eqref{Eq: SectionF_3}} into {\bl\eqref{Eq: Section3_2_14}}, $D_{\rm d}^{\left(2\right)} \phi \left(x'\right)$ is given by
\begin{equation}\label{Eq: SectionF_4}
D_{\rm d}^{\left(2\right)} \phi \left(x'\right) = \frac{\phi' \left(x' + d \right) - \phi' \left(x' - d \right)}{2d},
\end{equation}
since $\phi''' \left(x' \right) = 2x_0^3$. Substituting {\bl\eqref{Eq: SectionF_2}} into {\bl\eqref{Eq: SectionF_4}}, such that
\begin{equation}\label{Eq: SectionF_5}
D_{\rm d}^{\left(2\right)} \phi \left(x'\right) = \phi'' \left(x' \right).
\end{equation}
By setting $D_{\rm d}^{\left(1\right)} \phi \left(x'\right) = 0$ and $D_{\rm d}^{\left(2\right)} \phi \left(x'\right) = 0$, we have
\begin{equation}\label{Eq: SectionE_6}
x = - \frac{k}{4 x_0^3} \frac{1}{z} + \frac{k}{2z_{\rm F} x_0^3} + \underbrace {\frac{d^2 x_0^3}{3k}z}_{{\rm Part}\:6},
\end{equation}
where $z > 0$. Then, comparing Part 6 with the caustics in a continuous plane as in {\bl\eqref{Eq: SectionD_5}}, \textbf{\textit{Theorem 4}} is proved. \qed

\section{Proof of Proposition 6}

Based on the Nyquist sampling theorem, the phase difference between adjacent sampling points is less than or equal to $\pi$, such that
\begin{equation}\label{Eq: SectionG_1}
\max_{-\frac{L}{2} \le x' \le \frac{L}{2}} \left| \omega' \left(x' \right) \right|d \le \pi,
\end{equation}
where $\omega' \left(x' \right)$ is the first-order partial derivative of $x'$, while $\omega$ is set as in {\bl\eqref{Eq: Section3_1_11}}, such that
\begin{equation}\label{Eq: SectionG_2}
\omega' \left(x'\right) = x_0^3x'^2 - \frac{k}{z_{\rm F}}x' + \frac{kx_{\rm F}}{z_{\rm F}}.
\end{equation}
The formula {\bl\eqref{Eq: SectionG_2}} is a quadratic function of $x'$, while according to {\bl\eqref{Eq: Section3_2_7}}, $\left| \frac{k}{2 z_{\rm F} x_0^3} \right| \gtrsim \frac{L}{2}$. Consequently, we have
\begin{equation}\label{Eq: SectionG_3}
\begin{aligned}
&\max_{-\frac{L}{2} \le x' \le \frac{L}{2}} \left| \omega' \left(x' \right) \right| = \max \left|\omega' \left(\pm \frac{L}{2} \right)\right| \\ = &\max \frac{\left|z_{\rm F}x_0^3L^2 \pm 2kL + 4kx_{\rm F}\right|}{4z_{\rm F}}.
\end{aligned}
\end{equation}
Substituting {\bl\eqref{Eq: SectionG_3}} into {\bl\eqref{Eq: SectionG_1}}, \textbf{\textit{Proposition 6}} is proved. \qed

\section{Proof of Theorem 5}

According to {\bl\eqref{Eq: Section4_1_8}}, we have
\begin{equation}\label{Eq: SectionH_1}
\begin{aligned}
&P_{\rm A} |_{\eta \to 0} = \frac{P\lambda^2}{16\pi^2 N r_{\rm U}^2d^2} \left|d \sum\limits_{n = 1}^{N - \eta N} \exp \left[j\frac{x_0^3}{3}x_n^3\right] \right|^2 \\ \mathop \to \limits^{\left(\rm e \right)} &\frac{P\lambda^2 \left(1 - \eta \right)^2}{16\pi^2 r_{\rm U}^2 Nd^2} \left|\int_{-\infty}^{+\infty} {\exp \left[j\frac{x_0^3}{3} x'^3 \right]}dx' \right|^2,
\end{aligned}
\end{equation}
where (e) is derived by $N \to +\infty$ and $d \to 0$. Substituting {\bl\eqref{Eq: SectionH_1}} into {\bl\eqref{Eq: Section4_1_8}} and according to {\bl\eqref{Eq: Section4_1_4}}, {\bl\eqref{Eq: Section4_1_9}} and {\bl\eqref{Eq: Section4_1_10}} can be obtained. Consequently, for $x_0 \ne 0$, we have
\begin{equation}\label{Eq: SectionH_2}
\begin{aligned}
P_{\rm A} |_{\eta \to 0}^{N \to + \infty, d \to 0} = \frac{{\rm Ai}_{\rm max}^2}{N^2x_0^2 d^2} P_{\rm G} |_{\eta \to 0} \ll P_{\rm G} |_{\eta \to 0}.
\end{aligned}
\end{equation}
Furthermore, it is important to note that
\begin{equation}\label{Eq: SectionH_3}
\begin{aligned}
\frac{P_{\rm A} |_{N \to + \infty}^{\eta \to 1, d \to 0}}{P_{\rm G} |_{N \to + \infty}^{\eta \to 1, d \to 0}} = \frac{\left|C_0 \left(x_{\rm O}, z_{\rm O}, x_{\rm U}, z_{\rm U}\right) \right|^2 {\rm Ai}_{\rm max}^2}{x_0^2 d^2 \left| \phi_{\rm b} \left(x_{\rm U}, z_{\rm U}\right) \right|^2} |_{d \to 0} \to +\infty,
\end{aligned}
\end{equation}
where $\left| \phi_{\rm b} \left(x_{\rm U}, z_{\rm U}\right) \right| \to 0$. Hence, \textbf{\textit{Theorem 5}} is proved. \qed

\end{appendices}

\end{document}